\documentclass[aps,preprint,superscriptaddress,balancelastpage,twocolumn,10pt,showpacs,preprintnumbers,amsmath,amssymb,aps]{revtex4}

\usepackage{dcolumn}
\usepackage{graphicx}
\usepackage{hyperref}
\def\LL{\left\langle}	
\def\RR{\right\rangle}	
\def\LP{\left(}		
\def\RP{\right)}	
\def\LB{\left\{}	
\def\RB{\right\}}	
\def\PAR#1#2{ {\frac{\partial #1}{\partial #2}} }
\def\PARTWO#1#2{ {{\partial^2 #1}\over{\partial #2}^2} }

\newcommand{\BE}{\begin{displaymath}}
\newcommand{\EE}{\end{displaymath}}
\newcommand{\BNE}{\begin{equation}}
\newcommand{\ENE}{\end{equation}}
\newcommand{\BEA}{\begin{eqnarray}}
\newcommand{\EEA}{\nonumber\end{eqnarray}}

\newcommand{\etal}{{\emph {et al.}}}
\begin{document}

\title{The intrinsic strangeness and charm of the nucleon using improved staggered fermions}

\author{W. Freeman}
\affiliation{Department of Physics, George Washington University, 725 21st St. NW, Washington, DC 20052, USA}
\author{D. Toussaint}
\affiliation{Department of Physics, University of Arizona, 1118 East 4th St., Tucson, AZ 85721, USA}
\collaboration{MILC Collaboration}\noaffiliation

\date{\today}

\begin{abstract}
We calculate the intrinsic strangeness of the nucleon, $\LL N |\bar s s | N \RR\ -\ \LL 0 | \bar s s | 0 \RR$, using
the MILC library of improved staggered gauge configurations using the Asqtad and HISQ actions. Additionally, we 
present a preliminary calculation of the intrinsic charm of the nucleon using the HISQ action with dynamical charm. 
The calculation is done with a method which incorporates features of both commonly-used methods, the direct evaluation
of the three-point function and the application of the Feynman-Hellman theorem. We present an improvement on this method
that further reduces the statistical error, and check the result from this hybrid method against the other two methods 
and find that they are consistent.
The values for $\LL N |\bar s s | N \RR$ and $\LL N |\bar c c | N \RR$ found
here, together with perturbative results for heavy quarks, show that dark matter scattering through
Higgs-like exchange receives roughly equal contributions from all heavy quark flavors.
\end{abstract}

\maketitle

\section{Introduction}
\begin{figure}[htb]
         \includegraphics[width=3in]{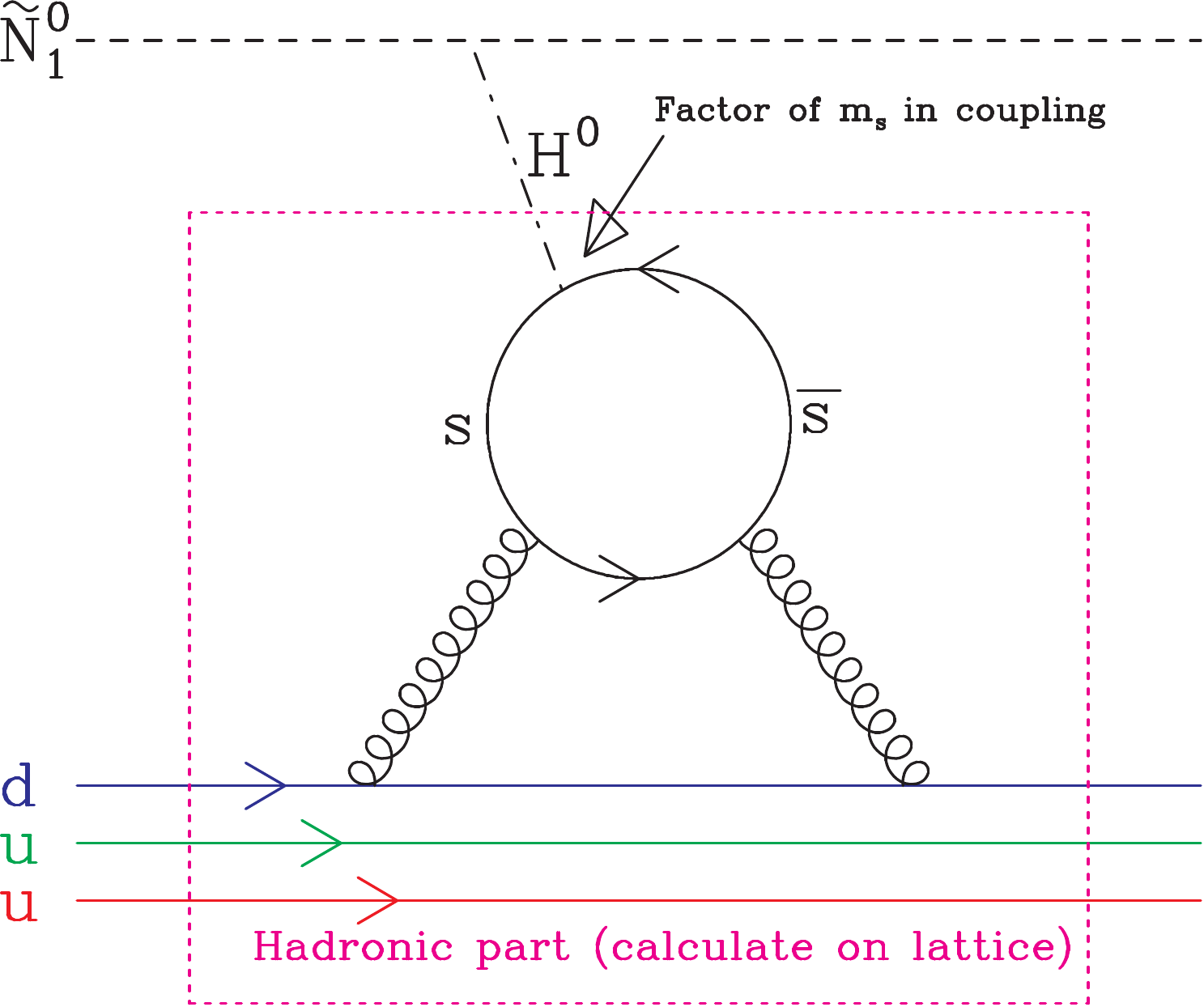}
\caption{Feynman diagram of an incoming neutralino interacting with a sea strange quark loop in the proton, mediated by a Higgs boson. The overall interaction amplitude depends on
the intrinsic strangeness of the nucleon which must be computed on the lattice.}
\label{fig-ssbar-loop}
\end{figure}

\begin{figure*}[htb]
\includegraphics[width=6in]{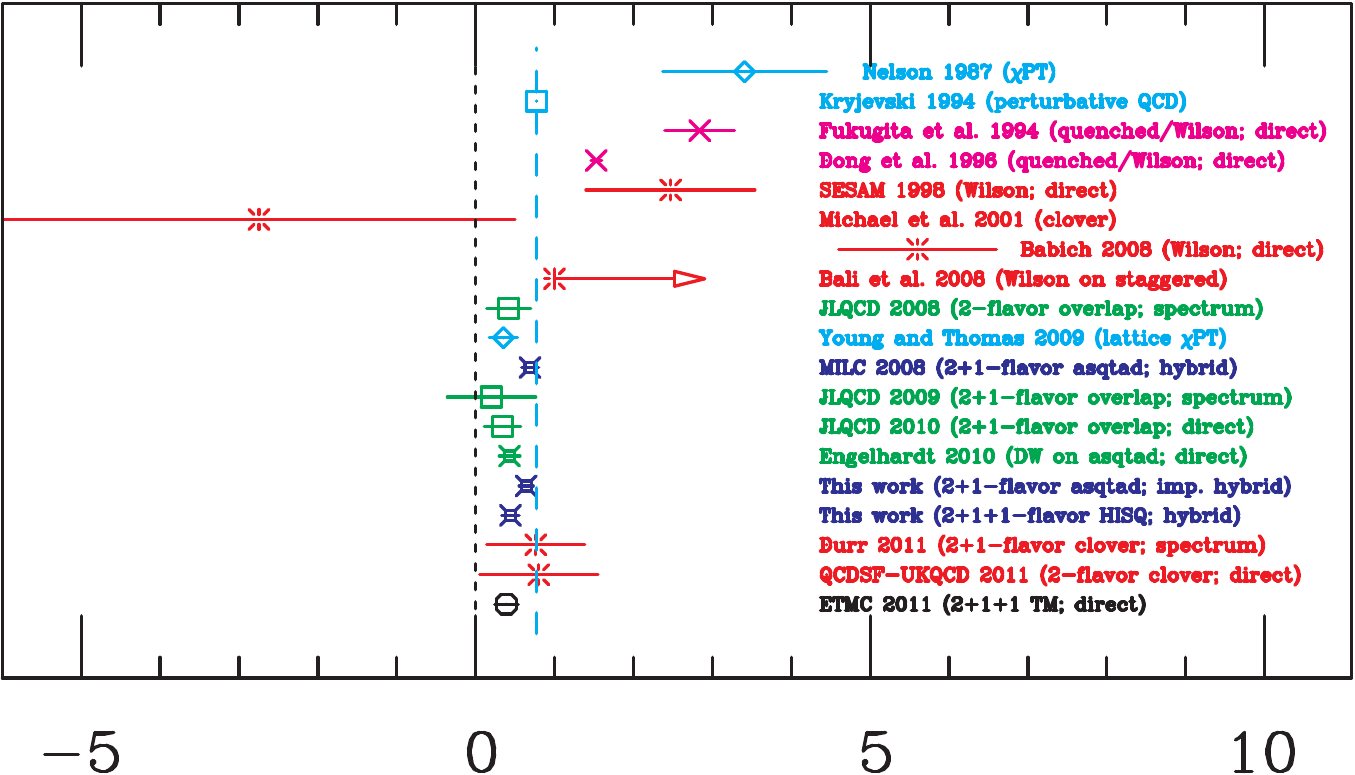}
\caption{A history of calculations of the nucleon strangeness. The ``natural scale'', given by the
perturbative QCD calculation\cite{KRYJEVSKI03} as discussed in \ref{sec-pert}, is shown as a 
vertical blue dashed line (color online).
\cite{NELSON87,FUKUGITA94,DONG96,SESAM98,MICHAEL01,BABICH08,BALI08,OHKI08,JLQCD09,JLQCD10,ENGELHARDT2010,DURR11,HORSLEY11,DINTER11,OURPRL,KRYJEVSKI03}
}
\label{fig-history}
\end{figure*}

The intrinsic strangeness of the nucleon, the matrix element
$\LL N | \int d^3x\ \bar s s | N \RR\ -\ \LL 0 |\int d^3x\ \bar s s | 0 \RR$ (often abbreviated
$\LL N | \bar s s | N \RR$, with the vacuum subtraction and volume integral omitted), has been
a quantity of significant interest in the past few years due to its relevance to WIMP-on-nucleon
scattering cross sections in many models\cite{BALTZ06,ELLIS08}. This scattering process is shown
in Fig. \ref{fig-ssbar-loop}. Early
$\chi$PT calculations suggested that its value might be quite large\cite{NELSON87}.
Early lattice calculations of its value suffered from large statistical errors and/or uncontrolled systematic effects
\cite{FUKUGITA94,DONG96,SESAM98,MICHAEL01,BALI08}, and the results are somewhat inconsistent. 
More recently, there have been a number of more modern calculations with better control of
systematic errors, notably the use of fermion actions retaining all or part of
the continuum chiral symmetries,  which are roughly in agreement\cite{OHKI08,JLQCD09,JLQCD10_POS,ENGELHARDT2010,DURR11,HORSLEY11,DINTER11},
including earlier work by us\cite{OURPRL} which is extended here. Fig. \ref{fig-history} shows a graphical depiction
of the history of calculations, mostly using lattice QCD, of the matrix element $\LL N | \bar s s | N\RR$. 

We present a refinement of this method, originally discussed in Ref. \cite{LAT2010PROCEEDING}, which
reduces statistical error while requiring, in principle, no additional computational effort. However, 
due to averaging of propagators and condensate measurements done by MILC in their original Asqtad analysis,
application of this method requires recomputation of propagators and the quark condensate. We thus only
use the refined technique on a subgroup of the available Asqtad ensembles.
Applying these methods to the MILC Asqtad data\cite{RMP} to calculate $\LL N | \bar s s | N \RR$, we obtain results
very similar to those in Ref. \cite{OURPRL}, but with smaller statistical and systematic errors. 

We also perform a direct evaluation of $\LL N | \bar s s | N \RR$ on those Asqtad ensembles where 
the propagators and condensate measurements have been rerun. This technique, while inferior on the MILC
dataset, has been used by a number of other calculations, for instance \cite{ENGELHARDT2010,JLQCD10_POS},
and is more theoretically straightforward; thus, it provides a useful cross-check with our ``hybrid method''
for $\LL N | \bar s s | N \RR$. The results from these methods are in excellent agreement.

We also apply this method to the newer HISQ gauge configurations to calculate both $\LL N | \bar s s | N \RR$
and $\LL N | \bar c c | N \RR$.
The result for the intrinsic strangeness is somewhat lower on the HISQ data, although it is not wildly different;
the result for the intrinsic charm has large errors but is consistent with a perturbative prediction.

\section{The meaning of the ``nucleon strangeness''}
The term ``intrinsic strangeness of the nucleon'' for this matrix element is somewhat deceptive. While in perturbation theory
the relevant diagram involves a sea strange quark loop (see Fig. \ref{fig-ssbar-loop}), this does not imply that the
presence of the valence light quarks in the nucleon somehow elicits strange quark loops
from the vacuum where none would exist otherwise. 
In fact, the meaning of the nucleon strangeness is almost exactly the opposite of this.
The vacuum strange quark loops, as part of the 
strange quark chiral condensate, pervade all space;
the presence of the valence quarks, in fact, 
partially {\it suppresses} the natural vacuum condensate. The ``strangeness of the nucleon''
is really the suppression of this vacuum behavior.

\begin{figure}[bht]
\includegraphics[width=2.5in]{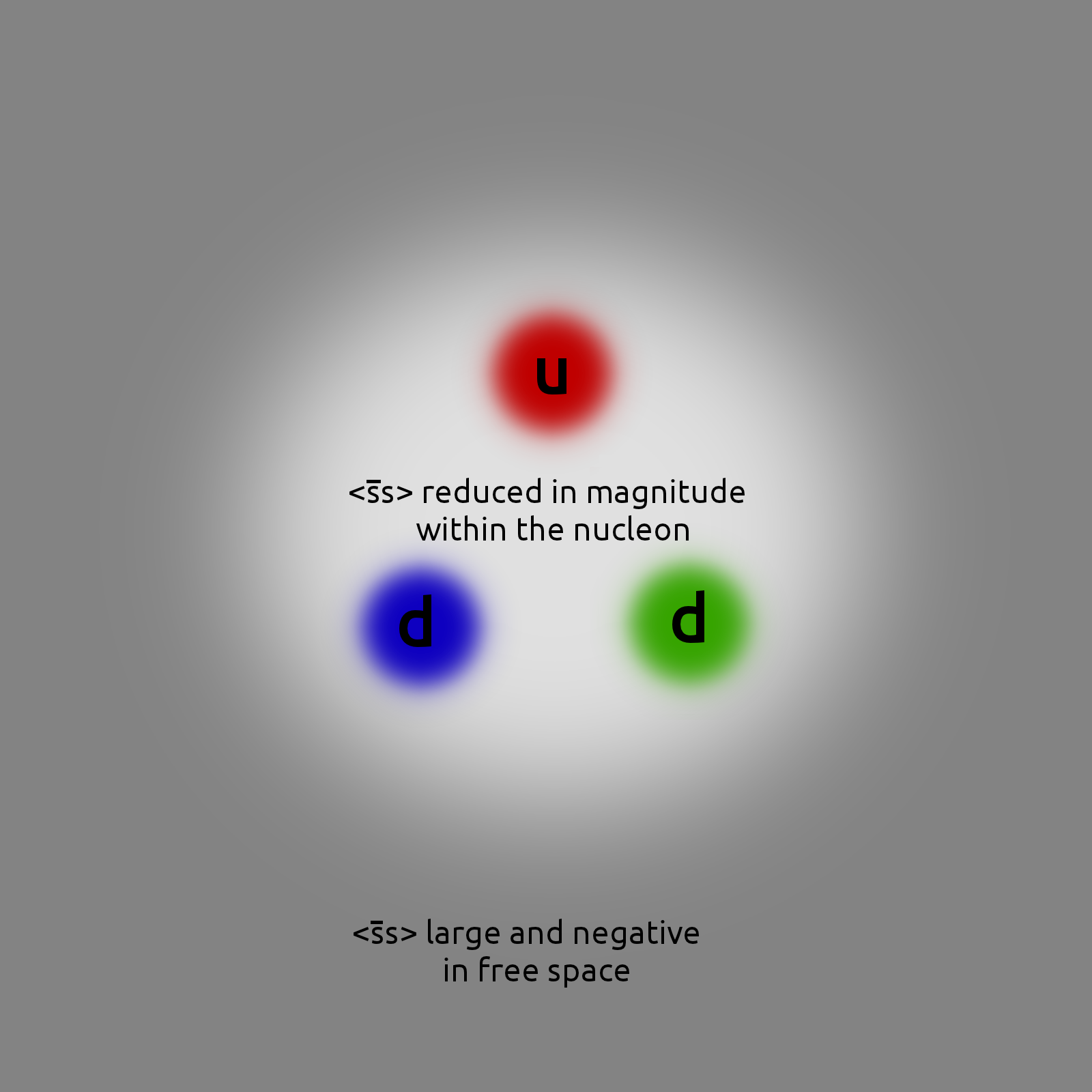}
\caption{A schematic illustration of the strangeness of the nucleon: the presence of the valence quarks creates a ``bubble'' in the vacuum chiral condensate.}
\label{fig-proton-bubble}
\end{figure}

Chiral symmetry for the strange quark is only an approximate symmetry; it is broken explicitly
by the mass term $m_s \bar s s$ in the QCD Lagrangian.
This explicit breaking determines the direction of the spontaneous breaking of chiral symmetry;
the symmetry is broken in the direction
that minimizes the action, leading to a negative expectation value for $\bar s s$ in the QCD vacuum,
with the usual sign convention that the mass term in the Euclidean Lagrangian is $+m_s \bar s s$.
In the presence of the valence quarks in the nucleon, the magnitude of the condensate is reduced. 
Since the ``strangeness of the nucleon'' is defined as $\LL N | \bar s s | N \RR - \LL 0 | \bar s s | 0 \RR$,
its natural sign is positive: both terms
are negative in sign, but the vacuum term is larger in magnitude. A schematic depiction of the nucleon's ``bubble''
in the vacuum strange
quark condensate is shown in Fig. \ref{fig-proton-bubble}.
The probability for an incident WIMP to scatter off of this bubble can be understood in the same
way as light scattering off of a bubble in a piece of glass: it is the change in the properties of the medium,
not the absolute presence or absence of those 
properties, that causes the scattering.

\bigskip

\subsection{Perturbative predictions}
\label{sec-pert}

The nucleon strangeness is an inherently nonperturbative quantity, since the strange quark is too light
to be treated properly in perturbation theory. For sufficiently heavy quarks, however, there is a
perturbative approach to calculate the $\LL N | \bar q q | N \RR - \LL 0 | \bar q q | 0 \RR$ matrix
element. This calculation is enlightening in two ways. First, it allows for a theoretical calculation
of the intrinsic charm of the nucleon. The intrinsic charm is similar in character to the intrinsic strangeness,
and can be calculated on the lattice in the same ways. This allows contact with the lattice result for the intrinsic charm.
Secondly, while there is no reason to expect it to be correct for the strange quark, the perturbative result
nonetheless sets a natural scale for the nucleon strangeness.  As we shall see, the value of
$m_q \LL N | \bar q q | N \RR$ is roughly constant for (perturbatively) heavy quarks. Lattice 
simulation must determine if the nucleon strangeness is substantially enhanced above this natural
scale (as suggested in Ref. \cite{NELSON87} and some early lattice studies) or not.


Beginning with work of Shifman, Vainstein and Zakharov \cite{SHIFMAN78} and continuing through
a four-loop perturbation theory calculation by Kryjevski\cite{KRYJEVSKI03}, the scalar condensate of a
heavy quark in the nucleon is
\BNE\label{PT_EQ1} \frac{m_q}{M_N} \langle N | \bar q q | N \rangle =
\frac{2}{33-2n_l} \LB 1 + C_1\alpha_s \ldots \RB \ \ \ .  \ENE
where $n_l$ is the number of quark flavors lighter than the heavy quark.
(The explicit perturbative corrections may be found in Ref.~\cite{KRYJEVSKI03}.)

The important point is that when a heavy quark mass $m_q \gg \Lambda, M_N$ is varied,
with the bare coupling constant held fixed, the nucleon mass is affected in the same
manner as the scale $\Lambda$ at which $\alpha_\Lambda$ runs to a particular value.
Since the running of the coupling
constant sets the scale for hadronic physics with light quarks,
this amounts to saying that varying a heavy quark mass has no effect on the physical value of $M_N$
or of any other low-energy lattice observable; rather, it amounts to an overall change in lattice scale setting.
(Recall that the quantity $M_N$ appearing in the Feynman-Hellman relation
$\PAR{M_N}{m_q} = \LL N | \bar q q | N \RR$ is the lattice nucleon mass,
not a nucleon mass in physical units, since the latter quantity is ambiguous and depends on the method
used to set the lattice scale: one could choose $M_N$ itself as the benchmark quantity to use in lattice
scale-setting, giving $\PAR{M_N}{m_q} \equiv 0$ under all circumstances!)

\begin{figure}[tb]
\includegraphics[width=3.0in]{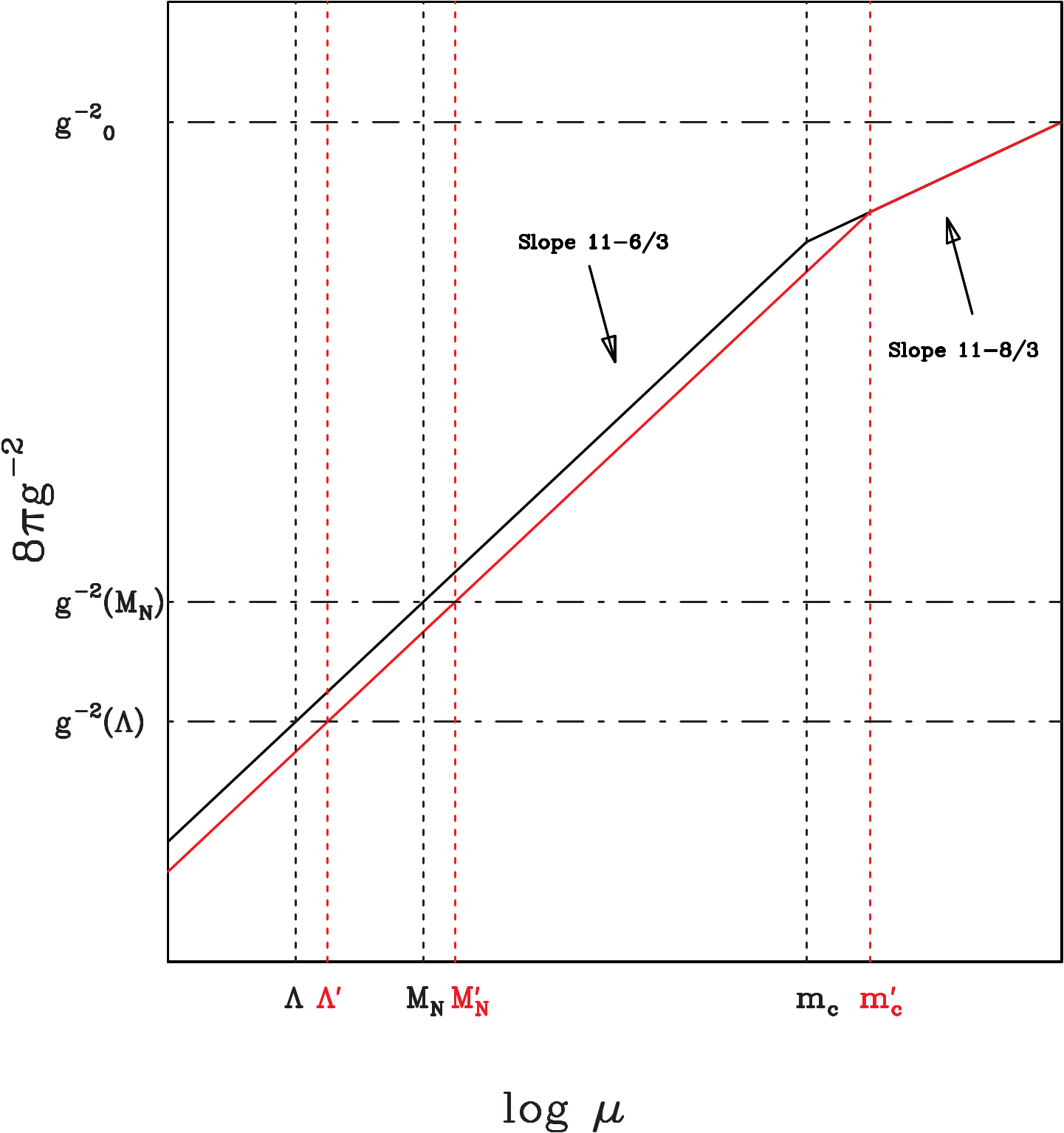}
\caption{A schematic depiction of the perturbative approach to $\PAR{M_N}{m_c}$ taken by Shifman and Kryjevski at leading order, inspired by a similar presentation by Kryjevski
in Ref. \cite{KRYJEVSKI03}. Changing the mass of a heavy quark changes the scale at which it freezes out and thus affects the running of the coupling constant,
affecting all low-mass scales equally.}
\label{fig-pert}
\end{figure}

Thus, $\PAR{M_N}{m_q}$ depends on the running of the coupling constant from its bare value $\alpha_0$
at scale $\mu_0$ to its value $\alpha_\Lambda$. At one-loop order, where the dependence of $g^{-2}$
on $\log \mu$ depends only on the number of light quark flavors, changing the charm quark mass from
$m_c$ to $m_c'$ only changes the scale at which the charm quark freezes out from the beta function, and the low-mass limit of
$\PAR{M_N}{m_c}$ can be calculated simply. (This is true for any other flavor of heavy
quark equally, of course.) This approach is shown graphically in Fig. \ref{fig-pert}.

While we would not trust a perturbative calculation at the strange quark mass, Eq.~\ref{PT_EQ1}
defines a natural scale for quark condensates, and lattice calculations are needed to see
if the strange quark content is in fact enhanced relative to this scale.

\section{Methods}

\subsection{The direct method}

A theoretically straightforward, but practically difficult, way to evaluate the nucleon strangeness is to
just evaluate the needed matrix element directly. One computes

\begin{align}
\LL N | \bar s s | N \RR = &\frac{\LL N^\dagger(0) N(T) \bar s s(T_1) \RR}{\LL N^{\dagger(0)} N(T) \RR} - \LL \bar s s \RR  
\label{direct-method-def}
\end{align}

\noindent where $T$ is the source-sink separation of the nucleon propagator, and $T_1$ is an intermediate time chosen
sufficiently far from both 0 and $T$ that the overlap with the excited states is small. (Here again volume integrals have been omitted.)

In the limit where $e^{-M_N n_t} << 1$ ({\it i.e.} states wrapping around the lattice are irrelevant), this can be written
as

\begin{widetext}
\begin{align}
&\LL N | \bar s s | N \RR =  
&\frac{\displaystyle \sum N^\dagger(T_0) N(T_0 + T) \bar s s(T_0 + T_1) + N^\dagger(T_0) N(T_0 - T) (-1)^T \bar s s(T_0 - T_1)}
{\displaystyle\sum N^\dagger(T_0) N(T_0 + T) + N^\dagger(T_0)N(T_0-T)(-1)^T}
&- \sum_{\textrm{configs}} \frac{1}{n_t}\sum_{t} \bar s s(t)
\label{direct-method-mc}
\end{align}
\end{widetext}

where $T_0$ are the source locations of the nucleon propagators, $T$ is their length, $T_1$ is an intermediate
length at which the strange quark condensate is measured, and the sums run over all gauge configurations and all source locations 
on each configuration. The factor of $(-1)^T$ comes from the fact that a backward-propagating
nucleon state in the staggered fermion formulation carries this factor.

This approach requires the evaluation of $ N^\dagger(0) N(T) $ at appropriate distances $T$,
along with the evaluation of the strange
quark condensate $\int d^3x\, \bar s s(x,t)$ on intermediate timeslice(s) $T_1$.
This is not part of the standard MILC analysis program,
which only records $ N^\dagger(0) N(T) $ averaged over all source timeslices (typically 8 per configuration),
 and only records the whole-lattice condensate $\int d^4x\ \bar s s(x,t)$ rather than separate values per timeslice.
Thus, without extra computer work to recompute correlators and condensates, this method cannot be used
on the MILC results.

\subsection{The spectrum method}

An alternate route to $\LL N | \bar s s | N \RR$ involves the Feynman-Hellman theorem, to equate $\LL N | \bar s s | N \RR$
with $\PAR{M_N}{m_s}$. This relation may be derived by differentiating the partition function with respect to $m_s$
and using the fact that $\LL N^\dagger(0) N(T) \RR = e^{-M_N T}$ for an ideal nucleon operator. This method has the disadvantage
that it requires multiple ensembles with different values of $m_s$ but all other lattice parameters held fixed, a condition
not met by the existing MILC data.

\subsection{The hybrid method}
\label{sec-hybrid}
The two traditional methods for calculating the nucleon strangeness
would be expensive for the MILC ensembles: the direct method would
require recomputation of propagators, while the spectrum method would require
different ensemble parameters. Thus, we use a third method,
originally presented in Ref. \cite{OURPRL}, which combines their advantages:
it can obtain a value for $\LL N | \bar s s | N \RR$
from a single lattice ensemble with arbitrary lattice parameters, using only the
whole-lattice average condensate and correlators
averaged over all source timeslices that are available.

The nucleon mass $M_N$ is obtained by a fit to the nucleon propagator $P(t)$ and as such can be thought of as a complicated
function of the propagator at different times: $M_N = f \LP P(t_1),P(t_2),P(t_3)...\RP $.
The crucial idea is that one can use the
chain rule for differentiation to rewrite the derivative:

\begin{equation}
{{\partial M_N}\over{\partial m_s}} = {{\partial M_N} \over {\partial P(t_1)}} {{\partial P(t_1)} \over {\partial m_s}}
+ {{\partial M_N}\over{\partial P(t_2)}} {{\partial P(t_2)} \over {\partial m_s}}
+ ... \end{equation}

The partial derivatives ${\partial M_N}\over{\partial P(t_i)}$ can be evaluated most simply by applying a small perturbation
to the nucleon propagator and examining the change in the fit result, while 
the second partial derivative ${\partial P(t_i)}\over{\partial m_s}$ can be evaluated by an application of the Feynman-Hellman
theorem in reverse to relate it to $\LL P(t_i) \overline {s} s \RR - \LL P(t_i) \RR \LL \overline {s} s \RR$.

The measurements of $\bar s s$ have been made using the commonly-used stochastic estimator technique;
see for instance Refs. \cite{DONG94,MATHURDONG94}. Typically, MILC has made such measurements as part
of lattice generation to monitor equilibration and simulation-time autocorrelations and
to use for subtracting zero temperature values in equation of state calculations.
On most ensembles, enough estimates are available that the fluctuation
of the stochastic estimator is a small part of the overall uncertainty.
On a few of the coarsest ($a \approx 0.12$ fm) ensembles,
we have run additional estimates of $\bar s s$ to ensure that the stochastic estimator
does not introduce any meaningful error. While other groups have found it expedient
to project out the low modes of the Dirac operator and calculate their contribution exactly,
we find that using these configurations it is sufficiently cheap and precise
to simply use repeated stochastic estimators on the entire space. 

Prior results obtained by applying this method to the MILC Asqtad gauge configurations
can be found in Ref. \cite{OURPRL}.

\subsection{The improved hybrid method}
\label{sec-imp-hybrid}
In the original hybrid method, a
major contribution to the statistical error in this calculation comes from fluctuations
in the quark condensate that have no physical correlation with the hadron propagator.
While the correlation between these fluctuations and the propagator averages to zero
in the limit of infinite statistics, with finite statistics it does not, and spurious correlations of this
sort are a major contributor to statistical error. 

Since there is no physical reason that fluctuations in the quark condensate
far from the propagator should be correlated with it, those fluctuations contribute only
noise and can be discarded without introducing bias;
in other words, we replace

\begin{align}
 \PAR{P(T)}{m_s} =& \LL P(T) \int d^3{\vec x}\, dt \, \bar s s(\vec x,t) \RR \nonumber \\ 
                 -& \LL P(T) \RR \LL \int \, d^3{\vec x} \,dt \, \bar s s(\vec x,t) \RR 
\end{align}

\noindent with

\begin{align}
\PAR{P(T)}{m_s} =& \LL P(T) \int \, d^3 \vec x \, \int_{t_1}^{t_2} dt \, \bar s s(\vec x,t) \RR \nonumber \\
                -& \LL P(t) \RR \LL
\int \, d^3 \vec x \, \int_{t_1}^{t_2} dt \, \bar s s(\vec x,t) \RR 
\label{hybrid-correlation}
\end{align}

\noindent where $t_1$ and $t_2$
are chosen sufficiently far from the propagation region so that they do not affect the final result.

The application of this method to the MILC Asqtad ensembles requires some extra computer time, since 
it requires separate values of $\bar s s$ on each timeslice and of the nucleon propagator for each
source location; these separate values were not saved originally due to a desire to economize on
storage. Due to the expense of this on the finest ensembles, we have only completed these measurements on the
$a \approx 0.12$ fm Asqtad ensembles and some of the $a \approx 0.09$ fm Asqtad ensembles.

\section{$\LL N | \bar s s | N \RR$ from the MILC Asqtad data}

\subsection {Validity tests of the improved hybrid method}
\label{sec-hybrid-validity}




The hybrid method is exact, but subject to the usual systematic errors that affect lattice
calculations: pollution from excited states, finite size effects, and lattice discretization
errors. The improved hybrid method, however, only considers the strange quark condensate on those timeslices 
that are meaningfully correlated
with the propagator, and is thus only valid if the correlation between the quark condensate and
the propagator falls off reasonably rapidly away from the propagation region.

To test this assumption, it is useful to calculate the contribution of each timeslice to the overall correlation between
the propagator
and the condensate, given by the expression

\begin{equation}
\LL P(T) \int d^3{\vec x} \, \bar s s(\vec x,t) \RR - \LL P(T) \RR \LL \int \, d^3{\vec x} \, \bar s s(\vec x,t) \RR 
\label{direct-correlation}
\end{equation}

\begin{figure}

\begin{center}

\vspace{-0.6in}
 \parbox[h][2in][c]{5.0in}{\sloppy $\null$
\hspace{-1.6in}
\includegraphics[width=2.2in]{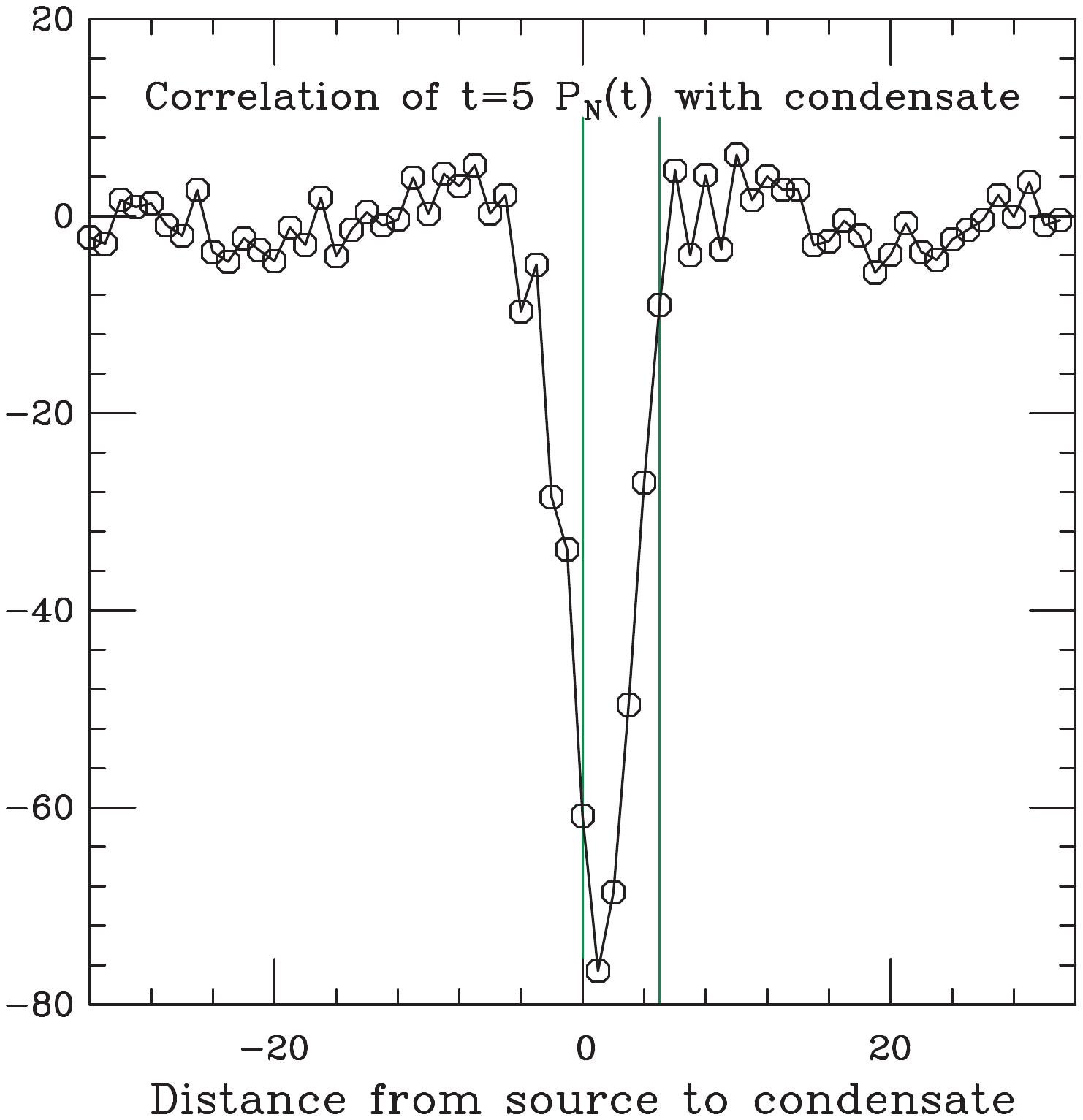}
\hspace{-0.6in}
\includegraphics[width=2.2in]{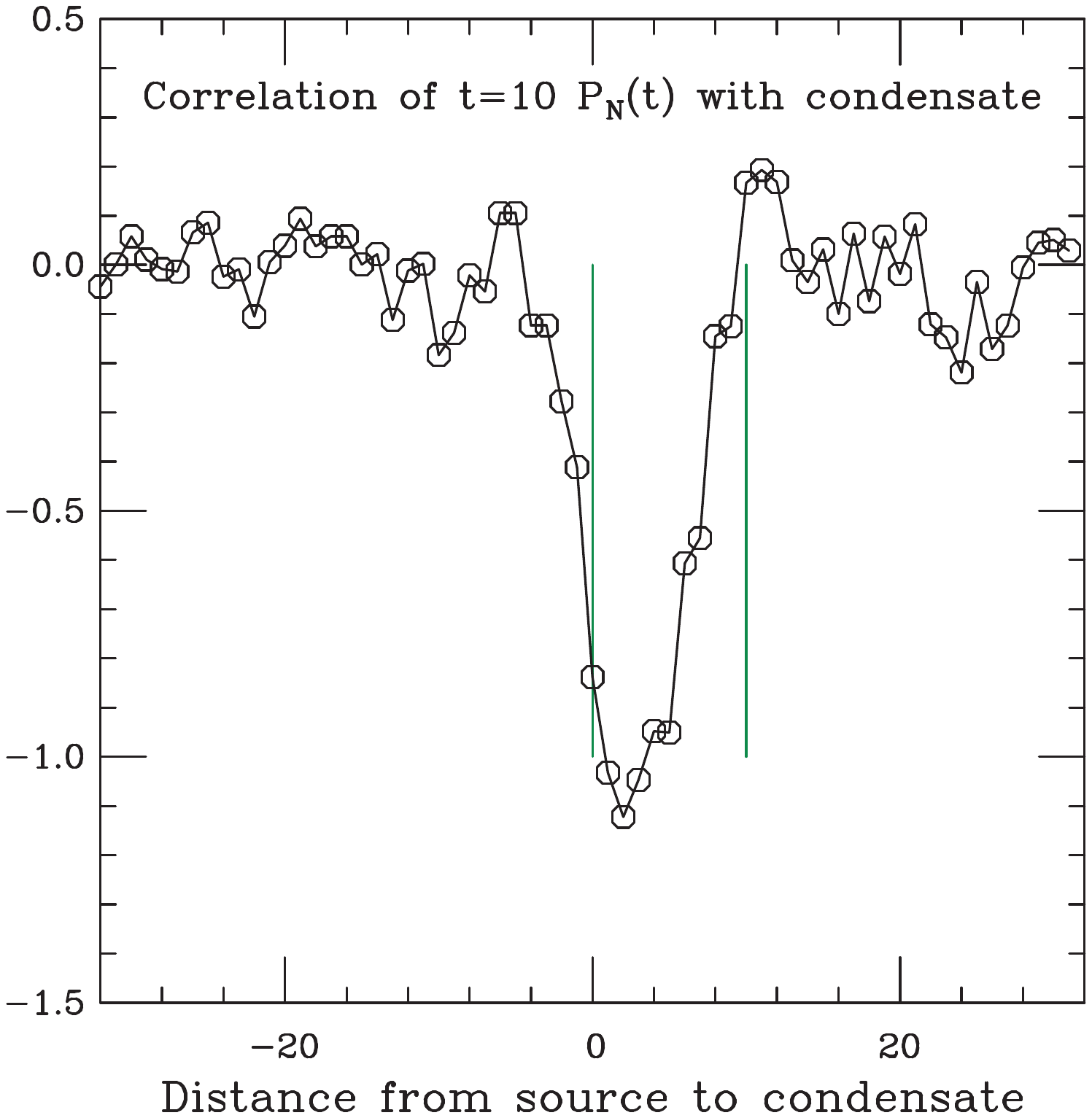}}
\end{center}
\caption{The correlation between nucleon propagators of length $5a$,  $10a$, and $15a$
with the strange quark condensate as a function of the distance from the propagator source
on one of the MILC $a=0.12$ fm ensembles. The source and sink location of the propagators are marked
as green vertical lines (color online).}

\label{sbs_prop_corr}
\end{figure}

(compare to Eq. \ref{hybrid-correlation}). This correlation is shown in Fig.~\ref{sbs_prop_corr}.

These data show that the condensate is indeed only meaningfully correlated with the propagator within the propagation region
and in a small window outside of it, confirming that this method is valid.

\subsection {Choice of padding length}
\label{sec-padding-length}

We must thus determine the appropriate size of the ``padding'' outside the propagation
region in which condensate measurements will be considered. If this padding length
is chosen to be too small, then the result will suffer from systematic error, as some of the
physical correlations are not being considered; if it is chosen too large, then the result
will be unnecessarily noisy and the full benefit from this method will not be realized.

\begin{figure}[t]
\vspace{-0.4in}
\hspace{-0.08in}
\center{\includegraphics[width=2.8in]{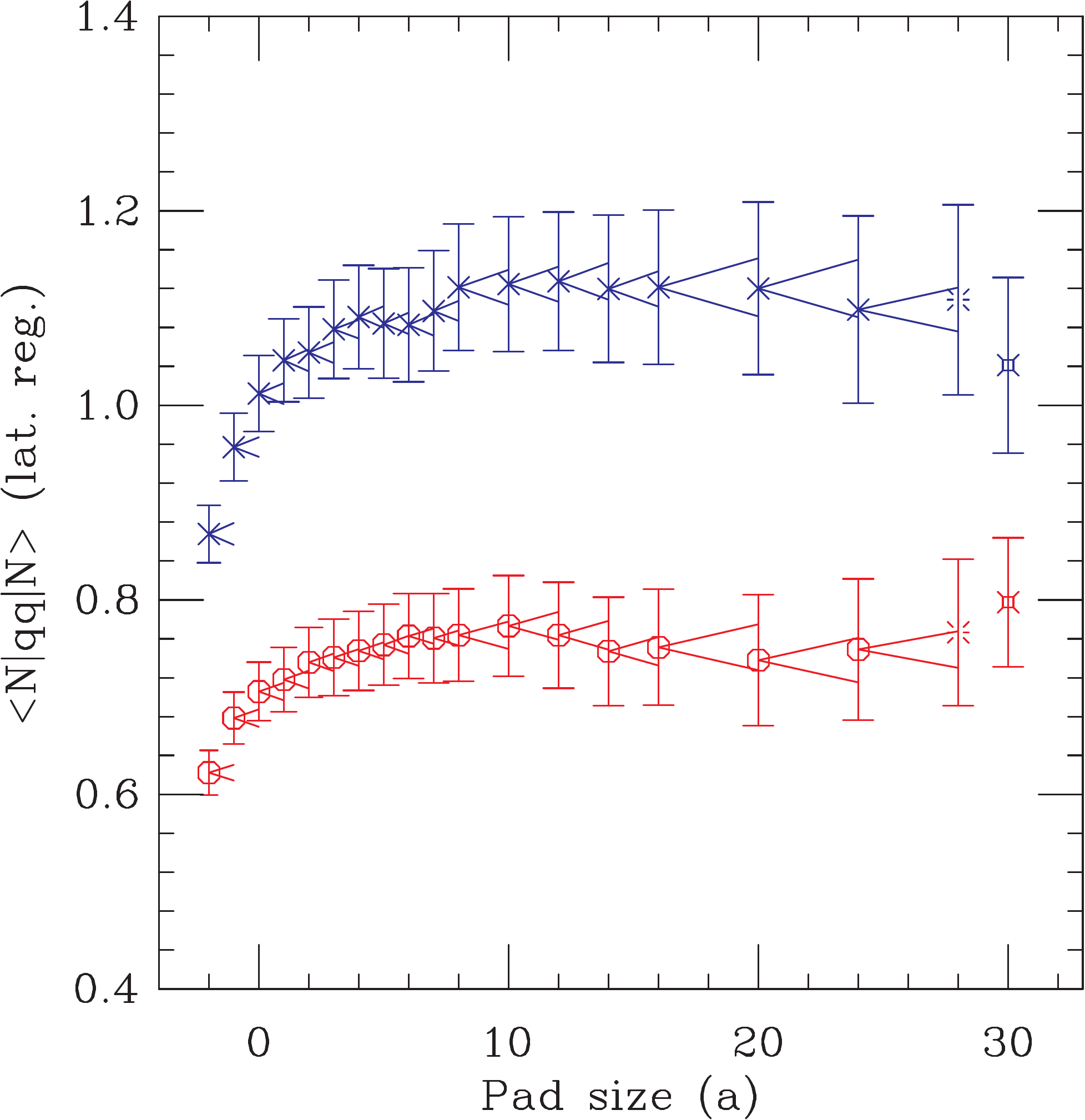}}

\caption{Results for $\LL N | \bar q q | N \RR$ using the improved hybrid method averaged
over the $a \approx 0.12$ fm ensembles on which the needed measurements are available
in the lattice regularization.  Each graph shows the values of $\LL N | \bar s s | N \RR$ (red circles)
and $\LL N | \bar u u | N \RR$ (blue crosses) at varying pad sizes (color online).
Negative pad sizes, with significant systematic bias, are included to make the trend of this bias plain.
The result using the unimproved method (i.e. the limit of large pad size) on the
newly-computed condensate and propagator measurements is shown as a starburst; the result using the
unimproved method on the old data is shown as a fancy square for comparison. The uncertainty in
the difference between adjacent values (obtained by jackknife) is shown as a horizontal carat
centered on the left point; if the right point lies within the bracketed range then the difference
between the two is less than one standard deviation.}

\label{fig-padsize-average}
\end{figure}

Fig.~\ref{sbs_prop_corr} suggests that the correlation between the condensate and the nucleon propagator
outside a window consisting of the propagation region and a window of width $\sim 5a$ on either side is
only noise. To see if only considering the condensate within this window improves the statistical error
without introducing significant bias, we consider the resulting value of
$\LL N | \bar s s | N \RR$ and $\LL N | \bar u u | N \RR$
for various widths of this padding on an average of the $a \approx 0.12$ fm ensembles.
We use the same choice of the minimum fit distance $T_{\rm{min}} = 5a$ as in Ref. \cite{OURPRL}.
The result is shown in Fig.~\ref{fig-padsize-average}. All errors have been determined by either
omit-10 or omit-20 jackknife; testing
indicated no meaningful dependence of error estimates on which jackknife blocksize was used.

The salient feature of these graphs is the desired one: a substantial reduction in statistical
error provided by the improved hybrid method. As expected, discarding physically unmeaningful
condensate measurements which contribute only noise reduces the error bars by quite a bit. 

Another notable feature is the difference between the results obtained from the old and new data
on the same ensemble using the unimproved method (the fancy square and burst points, respectively).
Part of this difference is due to the use of the stochastic estimator method to
determine $\bar s s$.  Indeed, by using the redone estimates of $\bar s s$ but the old propagators,
we see some difference between the results.  This effect, however, is not sufficient to explain
the entire difference. The old and new propagators themselves are slightly different due to the 
use of different source timeslices in the new data; when recomputing nucleon propagators, we used
eight equally-spaced source timeslices with the first located at a random offset from the lattice origin,
which will not necessarily correspond to the eight sources used originally.  This strongly suggests
that the MILC practice of using only eight source timeslices does not extract all available information
about the nucleon propagator from the lattice, and that the statistical errors
in $M_N$ and $\LL N | \bar s s | N \RR$ could be reduced further by the consideration of more source
locations (requiring, of course, more inversions to compute the propagators).

Examining \ref{fig-padsize-average}, we conclude that a conservative choice for padding size on
the $a \approx 0.12$ fm ensembles is $6a$, while an aggressive choice is $4a$. (The aggressive
choice was made in \cite{LAT2010PROCEEDING}; here we present data with the conservative choice,
which leads to essentially the same result but with slightly larger errors.) 
On the $a \approx 0.09$ fm ensemble, the equivalent padding size in physical units is $8a$.
We estimate the systematic bias due to this procedure to be no more than 1\%.

\subsection{Minimum fit distance and excited state pollution}

In Ref. \cite{OURPRL}, we chose minimum distances of $5a$, $7a$, and $10a$ on the
$a=0.12$, $0.09$, and $0.06$ fm ensembles respectively, and gave a rather conservative estimate of
the statistical error due to excited state pollution of $10\%$. These minimum distances, corresponding to a physical
distance of $\approx 0.6$ fm, were chosen to trade off statistical error
(worse at higher minimum distances) and the possibility of bias due to excited-state contamination. 
The largest obstacle to determining the appropriate minimum distance $T_{\rm {min}}$ and
the excited-state systematic error estimate was the overall large statistical error in
$\LL N | \bar q q | N \RR$, making it difficult to tell the difference between systematic bias and
statistical accident.  It is possible that the improved statistical errors from the improved hybrid
method may suggest a different choice of $T_{\rm {min}}$ or a different
estimate of the systematic error due to excited states.

\begin{figure*}[th]
\includegraphics[width=2in]{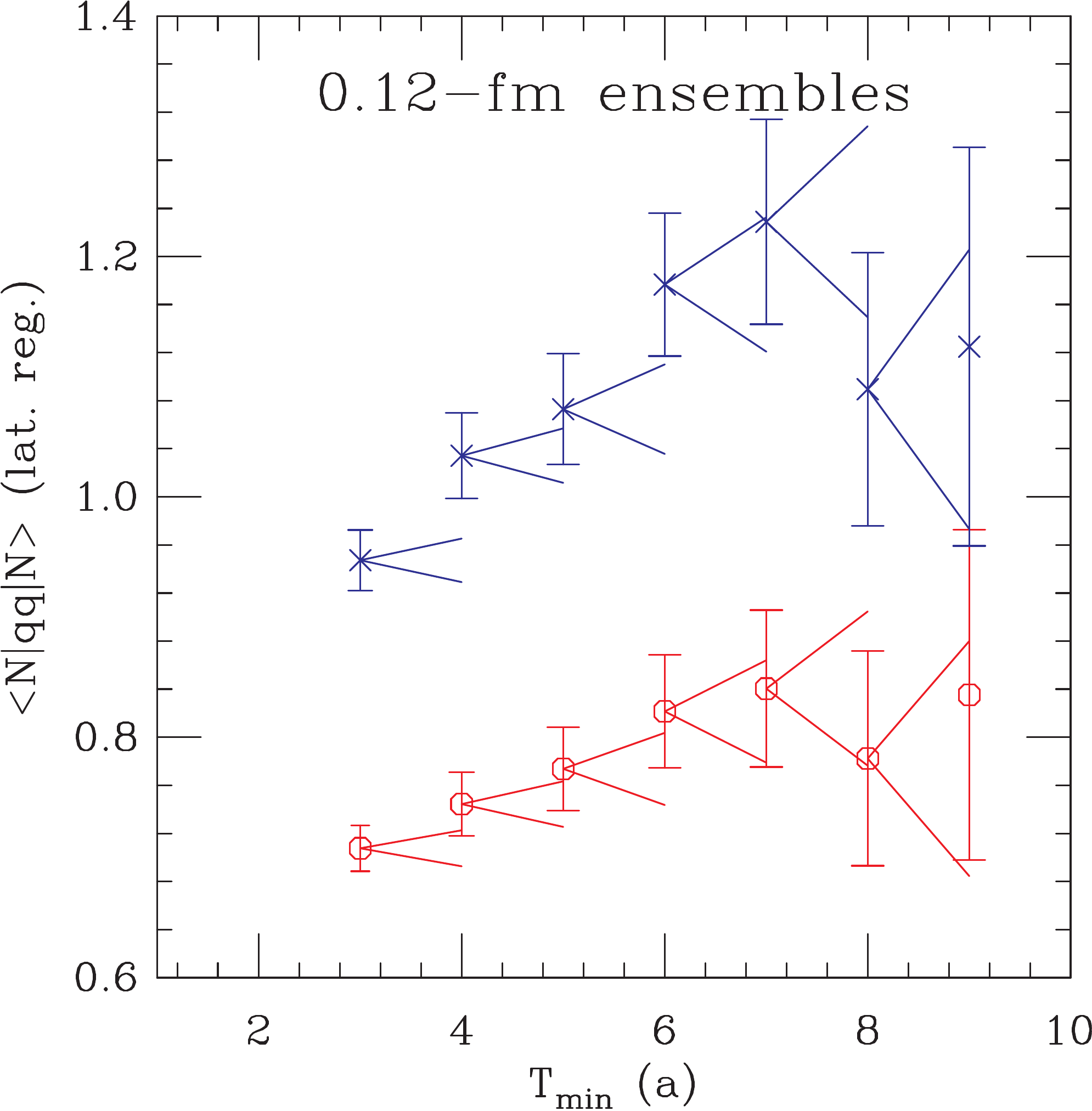}\hspace{0.25in}
\includegraphics[width=2in]{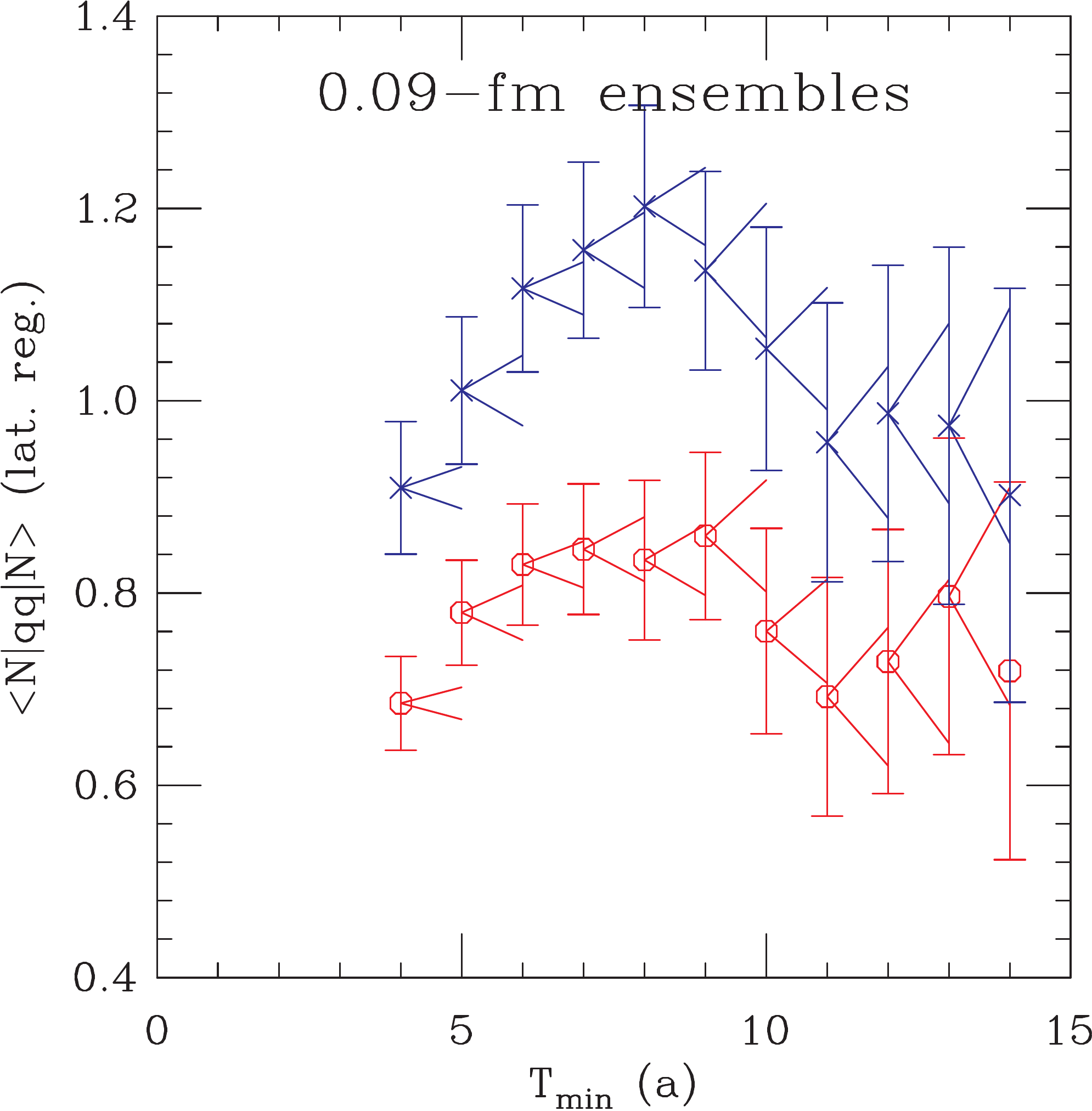}\hspace{0.25in}
\includegraphics[height=2.035in]{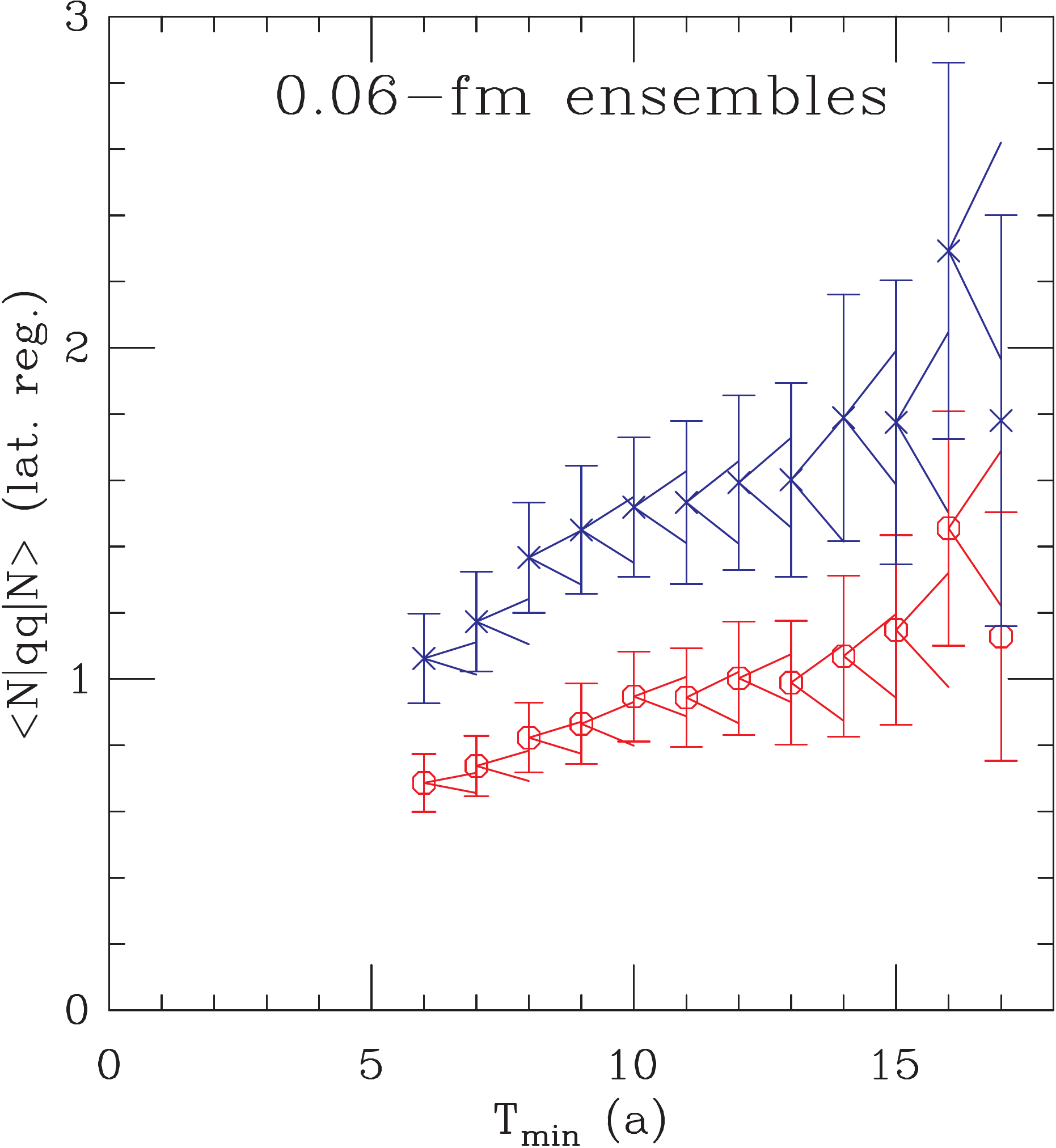}
\caption{$\LL N | \bar u u | N \RR$ (blue crosses) and $\LL N | \bar s s | N \RR$ (red circles) (color online) 
vs. $T_{\rm{min}}$ using the (improved) hybrid method. The three panels show the average values on the 0.12 fm, 0.09 fm, and 0.06 fm 
ensembles, respectively.
The uncertainty in the difference between adjacent values (obtained by jackknife) is shown as a horizontal carat centered on the left point.
}
\label{fig-dmin}
\end{figure*}

The left pane of Fig. \ref{fig-dmin} shows the $\LL N | \bar u u | N \RR$ and $\LL N | \bar s s | N \RR$ matrix elements (in the lattice regularization) averaged
over the five $a \approx 0.12$ ensembles on which the data for the improved method are available. We use the ``aggressive'' choice for padding size here ($4a$),
on the grounds that a little bit of systematic error due to padding will not obscure trends in $T_{\rm{min}}$, and that the more aggressive choice may allow
resolution of smaller systematic effects.

The difference between the average result on the coarse ensembles at $T_{\rm{min}}=5$ and $T_{\rm{min}}=6$ is
about one standard deviation: this could be due either to systematic bias or simply a statistical fluctuation.

However, the difference is substantially smaller than $1\sigma$ on the $a \approx 0.09$ and $a \approx 0.06$ fm ensembles,
suggesting that this $1\sigma$ difference is due to statistics rather than excited-state pollution.
The right pane of Fig. \ref{fig-dmin} shows the dependence of $\LL N | \bar u u | N \RR$ and $\LL N | \bar s s | N \RR$
on $T_{\rm{min}}$ for the $a \approx 0.09$ ensembles, in which no significant difference between $T_{\rm{min}}=7$
(chosen as the optimum minimum distance) and $T_{\rm{min}}=8$ is apparent.
The more precise results available with the improved method, thus, are still consistent with the previous
choices of $T_{\rm{min}}$.
Note that the aim in the choice of $T_{\rm{min}}$ is not to eliminate all systematic error from excited states,
but to achieve a balance between statistical error (larger at higher $T_{\rm{min}}$) and systematic errror 
(larger at lower $T_{\rm{min}}$).
The absolute size of this difference is consistent with a 5\% estimate for the systematic error due to excited states. 
Note that due to autocorrelations between the errors, an increasing trend is not necessarily indicative of a
systematic effect, and may just be a correlated statistical fluctuation in adjacent points. 

This lower estimate for the systematic error due to excited states can also be justified by examining other
information. The nucleon mass itself can be computed with much lower statistical error at higher $T_{\mathrm{min}}$;
mass fits at these higher minimum distances differ from those at $T_{\mathrm{min}} = 0.6$~fm
by only 1\% to 5\%. 
Na\"{i}vely applying the perturbative argument to the nucleon strangeness, the quantity
$\frac{m_s}{M_X} \LL N | \bar s s | N \RR$ is the same for any hadron $X$, including the
nucleon or the delta; thus, the fractional difference between the strangeness of the delta 
and the nucleon should be of the order of the fractional difference in their masses.
We also note that in the quark model the nucleon and delta differ in the spin of the
quarks, which we expect to be only indirectly related to the scalar strange quark content.
Thus, we may expect the systematic error in $\PAR{M_N}{m_s}$ due to
excited state pollution to be smaller than the 1\%-5\% range for the systematic error in the nucleon mass. 
(The 1\%-5\% range does not imply that this systematic shift varies greatly from ensemble to ensemble;
rather, at large $T_{\mathrm{min}}$ $M_N$ itself can only be determined to a few percent accuracy.)

It is possible to calculate a rough estimate of the fractional amount of excited-state pollution in a propagator
of any given length using a nucleon propagator fitting method that includes the lowest-lying
direct-parity excited state. (Note that due to the spin structure of staggered fermions,
the nucleon interpolating operator used here overlaps in part with the delta, which is this lowest ``excited state''.)
These methods give results for the amplitude and mass of the delta, allowing computation of the
ratio $\frac{A_\Delta e^{-M_\Delta T}}{A_N e^{-M_N T}}$ for any given propagator length $T$.
For the chosen minimum distances, this ratio is generally less than 0.1. Since the nucleon-delta mass splitting
is about 30\%, and we expect the nucleon-delta strangeness difference to be not much greater than 30\%,
this gives an estimate on the systematic shift due to delta pollution of 3\%.

\subsection{Renormalization}
\label{sec-renormalization}
The quantity $\PAR{M_N}{m_s}$ is renormalization scheme and scale dependent, since $m_s$ depends on the renormalization
scheme and scale but $M_N$ does not.
Thus, as a first step in analysis we convert the results on each ensemble 
to a consistent renormalization scheme, such as $\overline{\text {MS}}$ (2 GeV).
We used the Z-factors for converting from the Asqtad formulation to $\overline{\text {MS}}$ (2 GeV)
which were calculated by the HPQCD Collaboration
up to two-loop order in perturbation theory\cite{HPQCD2005}.

An alternative approach would be to work with the renormalization scheme
dependent quantity $m_s \LL N | \bar s s | N \RR$, or
$F_{T_s} \equiv \frac{m_s}{M_N} \LL N | \bar s s | N \RR$.

\subsection{Chiral extrapolation}
\label{sec-hybrid-chiral}
With one exception (with poor statistics), all of the MILC Asqtad gauge ensembles were performed with light quark masses
significantly larger than the physical value of $m_l$. This necessitates an extrapolation of any lattice measurement
to the physical light quark mass. 
Care must be taken in this extrapolation since the values of many quantities
curve sharply very near the physical point, and a na\"{i}ve polynomial extrapolation done on lattice data calculated
with a typical range of light quark masses $m_l > 10$ MeV will miss this curvature.

The statistics in this study
are not sufficiently strong to resolve subtle nonlinearities in the chiral form (or, equivalently, to determine higher-order
low-energy constants); they suffice only to do a linear extrapolation to the physical point. Thus, the presence of substantial curvature
for low light quark mass
(caused by $\frac{\partial^2 N}{\partial m_s\,\partial m_l}$ diverging as $m_l \rightarrow 0$ due to chiral logs) would cause
substantial difficulties in the chiral extrapolation.

The mass of the nucleon in terms of the quark masses and (mostly unknown) $\chi$PT coefficients has been calculated
up to order $m_q^2$ in chiral perturbation theory\cite{FRINKMEISSNER04}, with the explicit
form given in section 4 of this reference.
Differentiation with respect to $m_s$ gives the chiral fit form for $\PAR{M_N}{m_s}$.
This form can then be examined for ``dangerous'' terms whose derivatives with respect to $m_l$
diverge as $m_l \rightarrow 0$.

In this expansion there are no terms for which $\PAR{M_N}{m_s}$ diverges as $m_l \rightarrow 0$.
Another notable feature is the absence of a term $\approx m_s m_l \log(m_l)$.  (This
would be $\epsilon_{9,N}$ in the notation of Ref.~\cite{FRINKMEISSNER04}, and would give
a contribution to $\PAR{M_N}{m_s} \sim m_l \log(m_l)$, formally larger than a correction
linear in $m_l$. )
(We thank Ulf Meissner for correspondence on this point.)
Thus we can use a simple linear chiral extrapolation, of the form
\begin{equation}
\PAR{M_N}{m_s} = A + B m_l
\end{equation}
%

\subsection {Correcting for the strange quark mass}
MILC tried to run most of the Asqtad ensembles at the physical value of the strange quark mass.
However, since the lattice spacing is only determined {\it a posteriori} and since the physical strange quark mass itself
is not trivial to measure, the Asqtad ensembles were run at nonphysical strange quark masses. Since the aim of this 
work is to calculate the strangeness of the nucleon for physical strange quarks, an adjustment is necessary.
To correct for this and extrapolate the results to the physical value of $m_s$, we must determine the quantity
$\PAR{}{m_s} \LL N|\bar s s|N \RR = \PARTWO{M_N}{m_s}$.

$m_s$ has been most recently calculated by HPQCD on MILC lattices\cite{MASON06}, and updated by MILC
as part of a comprehensive fit of light meson masses
and decay constants to chiral perturbation theory \cite{BERNARD09} as 89.0(0.2)(1.6)(4.5)(0.1) MeV
in the $\overline{\mathrm{MS}}$(2 GeV) regularization scheme, where the errors are statistical,
miscellaneous systematic, perturbative renormalization, and electromagnetic, respectively.
For the purposes of this work we treat the value as 89 MeV exactly; the contribution of the uncertainty
in $m_s$ to the overall error in $\langle N | \bar s s | N \rangle$ is not significant, since
$m_s$ only enters this quantity indirectly.
(Note that when we compute the RNG invariant $m_s \langle N | \bar s s | N \rangle$, where $m_s$ appears
directly, the uncertainty in $m_s$ is very important.)

The values of $m_s$ used in the MILC ensembles do not differ by enough to determine this quantity in any meaningful way by simply performing
a fit. Thus, we must resort to a trick. The present method can also be used to calculate the equivalent light quark matrix element,
equivalent to $\PAR{M_N}{m_{l,sea}}$, in exactly the same way. This quantity, interpreted as the amount that the nucleon suppresses the light
quark condensate, has behavior qualitatively similar to the strangeness of the nucleon if $m_l$ is sufficiently large that the strong enhancement
near the chiral limit is not relevant; see Sec. \ref{sec-hybrid-chiral}. On those MILC ensembles with the heaviest light quarks (for instance, where $m_l = 0.4 m_s$),
we may treat the light sea quarks as ``lighter strange quarks'', and write  
\begin{equation}
\PAR{}{m_s} \LL N|\bar s s|N \RR = \frac{\LL N|\bar s s|N \RR - \LL N|\bar u u|N \RR}{m_s - m_l}.
\end{equation}
This approach has the advantage that the numerator of the right-hand side is the difference of two correlated quantities, so the error on their
difference (determined {\it via} the jackknife method, as usual) will be smaller than the na\"{i}ve sum of their errors in quadrature.

We choose to apply this technique to all ensembles with $m_l \geq 0.15 m_s$. The results are shown in Table~\ref{strange-mass-depend-table}. 
We use the improved hybrid method on the one of these ensembles where it is available.
Fitting a constant to these results, we obtain

\begin{equation}
\PAR{}{m_s} \LL N|\bar s s|N \RR = -0.00331(28) \rm{MeV}^{-1}.
\end{equation}

This value has be used to extrapolate measured values of $\LL N|\bar s s|N \RR$ to the correct strange quark mass. The error on the slope of this extrapolation
has been incorporated into the overall statistical error.

 \begin{table}
\centering
\begin{tabular}{|c | c | c c | c |}
\hline
$\beta$ & a (nominal, fm) & $a m_l$ & $a m_s$ & $\PAR{}{m_s} \LL N|\bar s s|N \RR (\textrm{MeV}^{-1})$ \\
\hline
6.81 & 0.12 & 0.30 & 0.50 &  -0.0033(21)  \\
6.79 & 0.12 & 0.20 & 0.50 &  -0.0021(6)   \\
6.79 & 0.12 & 0.20 & 0.50 &  -0.0030(3)   (improved)\\
7.10 & 0.09 & 0.093& 0.31 &  -0.0067(19)   \\
7.11 & 0.09 & 0.124& 0.31 &  -0.0046(8)   \\
7.48 & 0.06 & 0.72 & 0.18 &  -0.0046(24)  \\
\hline
\multicolumn{4}{|c|}{Average} & -0.00331(28) \\
\hline
\end{tabular}
\caption[Results for $\frac{\LL N|\bar s s|N \RR - \LL N|\bar u u|N \RR}{m_s - m_l}$ on ensembles with
heavy light quarks]{Results for $\frac{\LL N|\bar s s|N \RR - \LL N|\bar u u|N \RR}{m_s - m_l}$ on ensembles with
heavy light quarks, presumed to be a good estimator of the dependence of the strangeness of the nucleon
on strange quark mass, $\PAR{}{m_s} \LL N|\bar s s|N \RR$. The results are given in the
$\overline{\mathrm{MS}}$(2 Gev) renormalization scheme. The result is also given for the $a \approx 0.12$ fm
ensemble on which the improved method has been run. Errors are obtained {\it via} jackknife of the difference in the proper way.
The result of fitting a constant to these values (using the improved method for the ensemble where it is available) is shown;
this fit has $\chi^2/\rm{d.o.f} = 7.76/4$.}

\label{strange-mass-depend-table}
\end{table}

\subsection {Continuum extrapolation}
\label{sec-continuum}
The leading-order discretization errors in the Asqtad action are $\mathcal{O}(a^2)$, so
a continuum extrapolation can be performed most simply by adding a term $C a^2$ to the
fit form used in the chiral extrapolation, and fitting to the form

\begin{equation}
\PAR{M_N}{m_s} = A + B m_l + C a^2.
\end{equation}

However, due to the large statistical error in this work ($\sim$ 10\%), particularly for
the $a \approx 0.06$ fm and $a \approx 0.09 fm$ gauge ensembles, it is difficult to
perform a well-controlled continuum extrapolation. Doing the fit given above na\"{\i}vely
would potentially lead to large uncertainty in $C$ and thus a large uncertainty in the continuum limit.
There is no reason to expect the dependence of $\PAR{M_N}{m_s}$ on the lattice spacing to be significantly
larger than other similar hadronic quantities, however, so it is appropriate to constrain the parameter $C$
with a Bayesian prior. In prior MILC Asqtad spectroscopy, the continuum extrapolations of the $\rho$, nucleon, and $\Omega^-$ differ by 4\%, 10\%, and 9\%,
respectively, from the values measured on the $a \approx 0.12$ fm ensembles,
so we use a Gaussian prior on $C$ centered at zero
and with a width corresponding to a shift of 10\% at $a=0.12$ fm.

\subsection{Error budget and result}

\begin{figure}[ht]
\center{\includegraphics[width=3.0in]{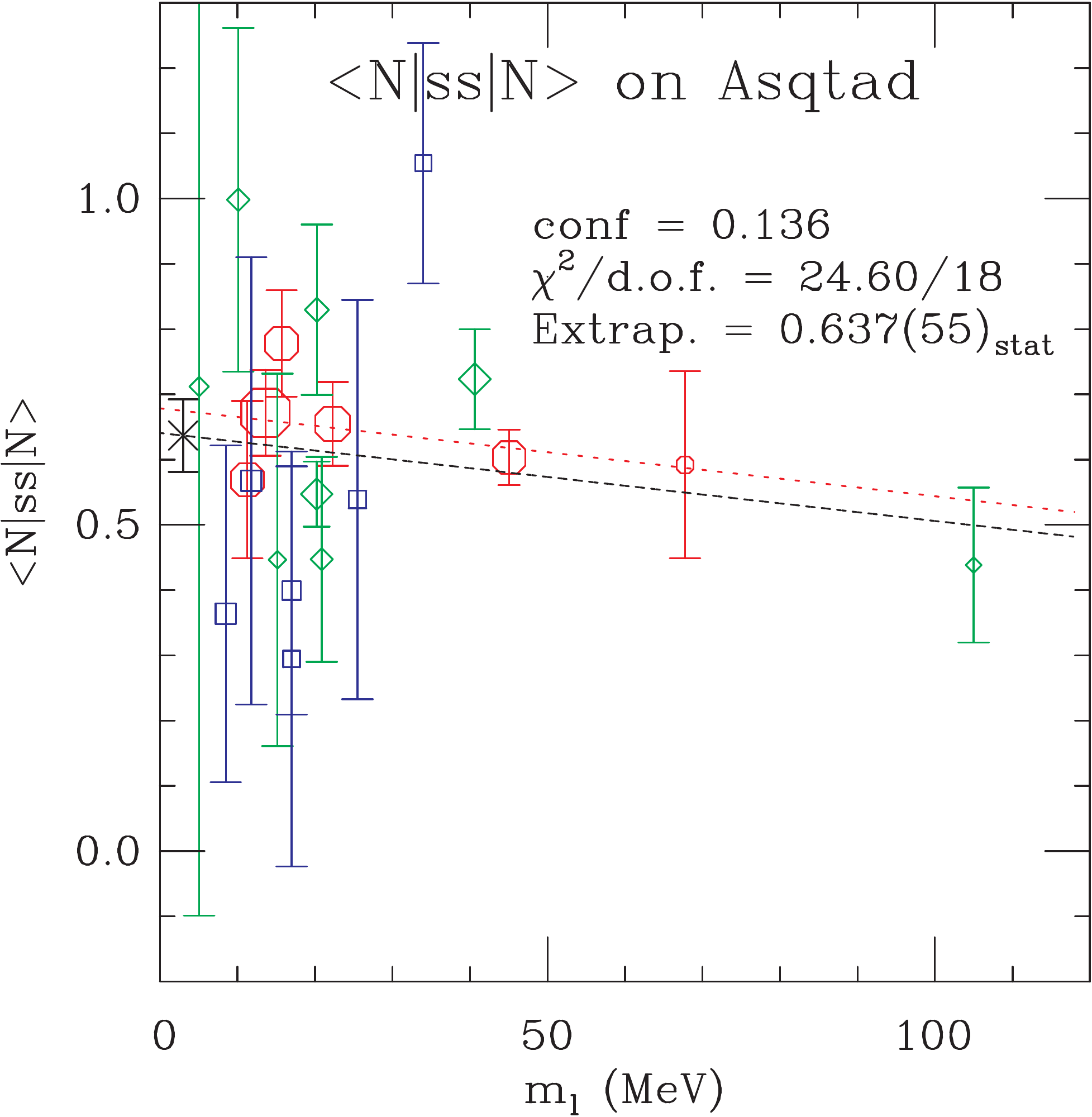}} 
\caption[$\LL N |\bar s s| N \RR$ on the Asqtad ensembles using data from the improved hybrid method]{$\LL N |\bar s s| N \RR$ on the Asqtad ensembles using data from the improved hybrid method where it is available, in the $\mathrm{\overline{MS}}$ (2 GeV) regularization scheme,
 adjusted to the correct value of $m_s$, along with the linear chiral and continuum fit.
                 Data from the 0.12, 0.09, and 0.06 fm ensembles are shown by red octagons, green diamonds, and blue squares, respectively.
                 The black line is the chiral fit in the continuum limit; the chiral fit at $a \approx 0.12$ fm is shown by the red dashed line (color online).
                 Symbol area is proportional to the number of lattices in the gauge ensemble. The black point
                 marked by a cross is the evaluation at the physical point, along with the combined statistical and
                 continuum-extrapolation error.}
\label{fig-ssbar-imp-linear}
\end{figure}

The value of $\LL N | \bar s s | N \RR$ on each ensemble considered, along with the
fit described in the preceding sections and its evaluation at the physical point, is shown in Fig. \ref{fig-ssbar-imp-linear}.

There are other significant systematic errors. The fit form linear in $m_l$ is of course an approximation, so there is an error
due to missing higher-order terms in chiral perturbation theory. The effect of these higher-order terms cannot be reliably determined
from the data available. If the chiral fit is modified to include a quadratic term, then the central value changes significantly, but
the coefficient of the quadratic term is rather poorly determined and the curvature in the fit is rather extreme; it is unlikely that
this curvature represents an actual nonlinearity in the chiral extrapolation, given the analysis of the chiral perturbation theory form
given in Sec \ref{sec-hybrid-chiral}. To estimate the
systematic error from exclusion of these higher-order terms, we use the case of chiral fits to $M_N$ to estimate the size of the effect
from these higher-order terms. If the value of
$M_N$ is fit to constant-plus-linear and evaluated at the physical point, the result differs from the result obtained by fitting
to two extra orders in $M_\pi$ by 7\%; we thus take 7\% as an estimate of the systematic error in $\LL N | \bar s s | N \RR$ due
to higher-orders in $\chi$PT. This may be an overly-conservative estimate; it is entirely possible that the mass of the nucleon is
more sensitive to the masses of the valence quarks than is the nucleon's suppression of the strange quark condensate.

Finite-size effects in general are expected to be small on the MILC Asqtad ensembles, since they have relatively large physical volumes. However,
this quantity may be especially sensitive to finite-volume effects. Recall that the physical basis for the nucleon strangeness is that the presence
of the nucleon's valence quarks and the glue field holding them together carves out a ``bubble'' in the QCD vacuum with different characteristics,
including the suppression of the strange chiral condensate; since the size of the measured effect is directly related to the
size of this region, which could extend over the nucleon and its surrounding pion cloud, it is potentially especially sensitive to small volumes.
In general, it is not possible to directly estimate the size of the finite-volume effects. However, in one case, there are two MILC Asqtad ensembles
with the same lattice parameters but different physical volumes. One volume corresponds to the volumes used on other ensembles, while one has a
spatial extent 40\% greater. The measured values of $M_N$ on these ensembles differ by 1\%; since it is possible that the nucleon strangeness
is more sensitive to finite volume effects, we conservatively estimate the systematic error due to finite volume corrections as 3\%. A more direct
test of finite size effects is available on the HISQ ensembles, where three ensembles with identical lattice parameters except for the volume are available.

Finally, there is an uncertainty in the values of $Z_m$ used to convert the result from the lattice regularization to the $\mathrm{\overline{MS}}$ (2 GeV) scheme
of 4\% \cite{MASON06}. The result can be presented in the renormalization-invariant form $m_s \LL N | \bar s s | N \RR$, the strange quark sigma term, to eliminate
this error. However, in this case there is a systematic error of nearly the same size coming from uncertainty in the physical value of $m_s$ and from uncertainty in
lattice scale-setting.

This gives a result, extrapolated to the physical point, of $\LL N | \bar s s | N \RR = 0.637(55)_{\rm{stat}}(74)_{\rm{sys}}$.
The error budget is summarized in Table \ref{table-error-budget}. The improvement in the statistical error over that reported in
\cite{OURPRL} is not as large as was hoped (note that the improved hybrid method reduces statistical error by roughly half); this
is due to the lack of improved hybrid results on the finer ensembles due to the expense of recomputing propagators. Thus, the 
largest remaining contributor to the statistical error is uncertainty in the continuum approximation.

\begin{table}
\center{  \begin{tabular} { | c | c | }
  \hline
  Source & Error\\
  \hline
  Statistical & 0.05 \\
  Improved method  & 0.007 \\
  Higher order $\chi$PT & 0.05 \\
  Excited states & 0.03 \\
  Finite volume & 0.02 \\
  Renormalization & 0.03 \\
  \hline 
  \end{tabular}}
  \caption{Error budget for the measurement of $\LL N | \bar s s | N \RR$ using the (improved) hybrid method on the Asqtad data.}
  \label{table-error-budget}
\end{table}

\subsection{Validation from the direct method}

The additional measurements made to apply the improved hybrid method are precisely the measurements needed to apply
the direct method. Thus, all of the measurements
required to implement Eq. \ref{direct-method-mc} have already been made on the coarsest ($a \approx 0.12 \textrm{fm}$) ensembles.

We do not expect the direct method to be competitive with the improved hybrid method for an accurate determination of $\LL N | \bar s s | N \RR$
on these ensembles. It considers less information (fewer $\bar s s$ condensate measurements and fewer propagator lengths), and does not explicitly
consider the effect of excited states in the nucleon propagator. In particular, it does not consider the alternating-parity state which appears in
propagators using staggered fermions. However, it is the preferred method of many other groups calculating the nucleon strangeness, and thus it is appropriate
to apply it to the MILC Asqtad data where possible to provide a comparison between the methods. Since we are using it only as a comparison to the
(improved) hybrid method and not as a standalone calculation of the nucleon strangeness, we do not convert to the $\overline{ \rm{MS}}$ (2 GeV) regularization scheme
or apply corrections for the strange quark mass.

The direct method requires two choices: the length of the propagator ($T$) and the location within that propagator
($T_1$, measured here from the source)
where the condensate is evaluated. As usual, there is a systematic/statistical error tradeoff. If the condensate
measurement is chosen too close to either end of the propagator, then there is the possibility of contamination by
excited states; only far from the source and sink does the propagating state approximate a pure nucleon. If the propagator is
chosen to be too short, then nowhere between source and sink is this condition true, and any result will potentially
suffer from large systematic error. However, the signal-to-noise ratio of the propagator declines exponentially with
increasing distance, so choosing longer propagators leads to larger statistical error.

\begin{figure}

\hspace{-.2in}        \includegraphics[width=2.9in]{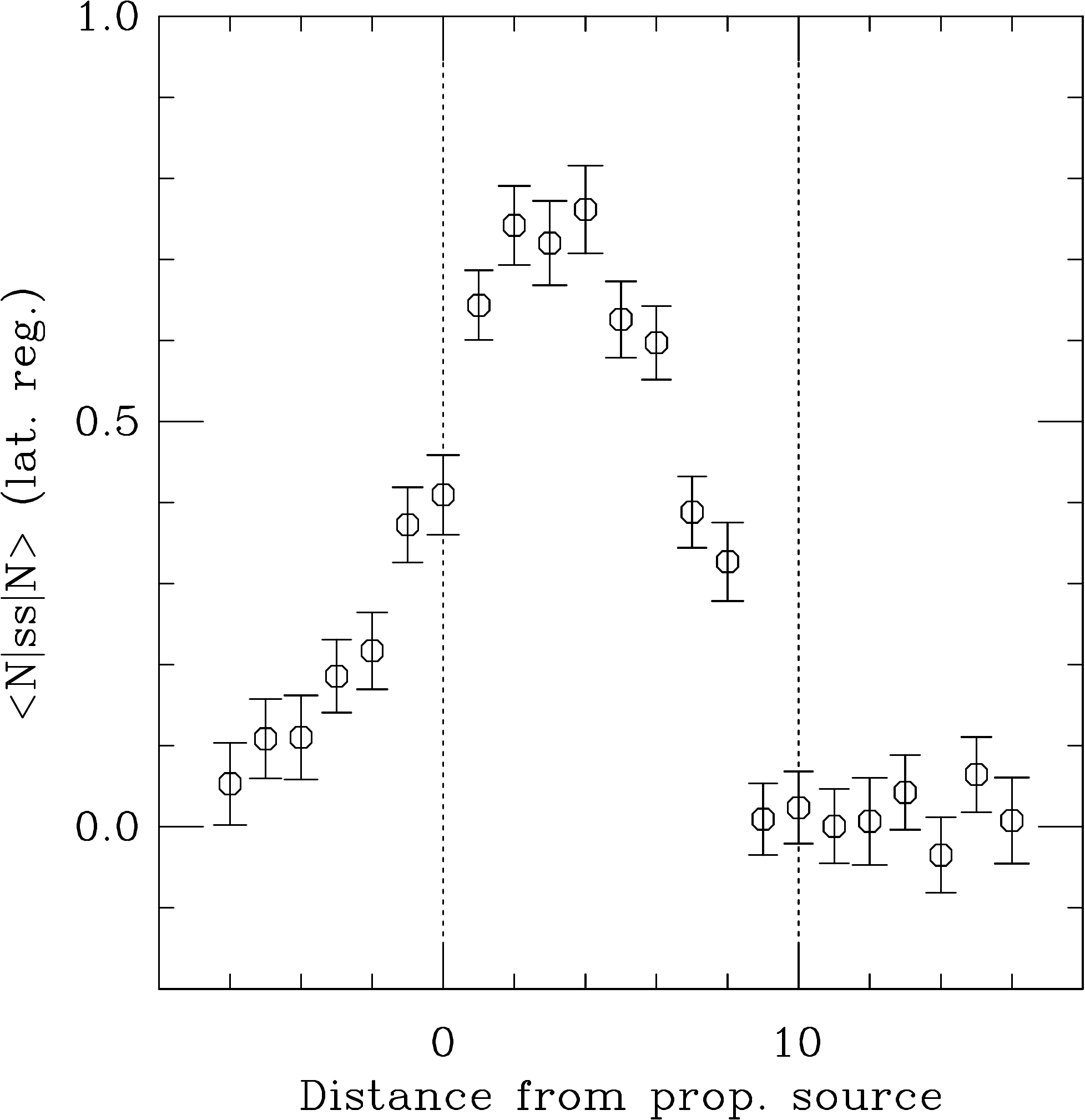}

\caption[$\LL N | \bar s s | N \RR$ using the direct method with a propagator length of $10a$]{The intrinsic strangeness of the nucleon on the MILC Asqtad ensembles using the direct method with a propagator length
of $10a$, as a function of $\bar s s$ insertion time $T_1$.
Vertical dashed lines represent the sink for the propagator,
while the horizontal lines show the average using the improved hybrid method with the ``conservative'' padding choice.}
\label{fig-direct-10}

\end{figure}

Fig. \ref{fig-direct-10} shows the result for $\LL N | \bar s s| N \RR$ as a function of the location of the $\bar s s$ measurement
(the value $T_1$ in Eq. \ref{direct-method-mc}) for a propagator length of $10a$, averaged over the five $a \approx 0.12$ ensembles where
the needed measurements have been made. As expected, the result depends strongly on the choice of $T_1$; the most accurate
values will be located far from both source and sink. While the length of the
propagator is too short to definitively say that a plateau exists, there is
certainly the suggestion of one in the region $2a \le T_1 \le 5a$.
This plateau agrees well with the calculation using the improved hybrid method, reinforcing the validity of that result.

Another very notable feature of this figure is the strong asymmetry between source and sink ends; the result is nearly to its plateau
value at the source but consistent with zero at the sink. This is due to the asymmetry of the operators used by MILC to calculate
the nucleon two-point function (Coulomb wall source, point sink).
Similar asymmetry is notable in Fig. \ref{sbs_prop_corr}, which hints that it would be possibly useful to explore an asymmetric
padding window in the improved hybrid method. This effect was also noted by the JLQCD collaboration in their application of the
direct method\cite{JLQCD10} using asymmetric operators.

\begin{figure}[h]

        \includegraphics[width=3.2in]{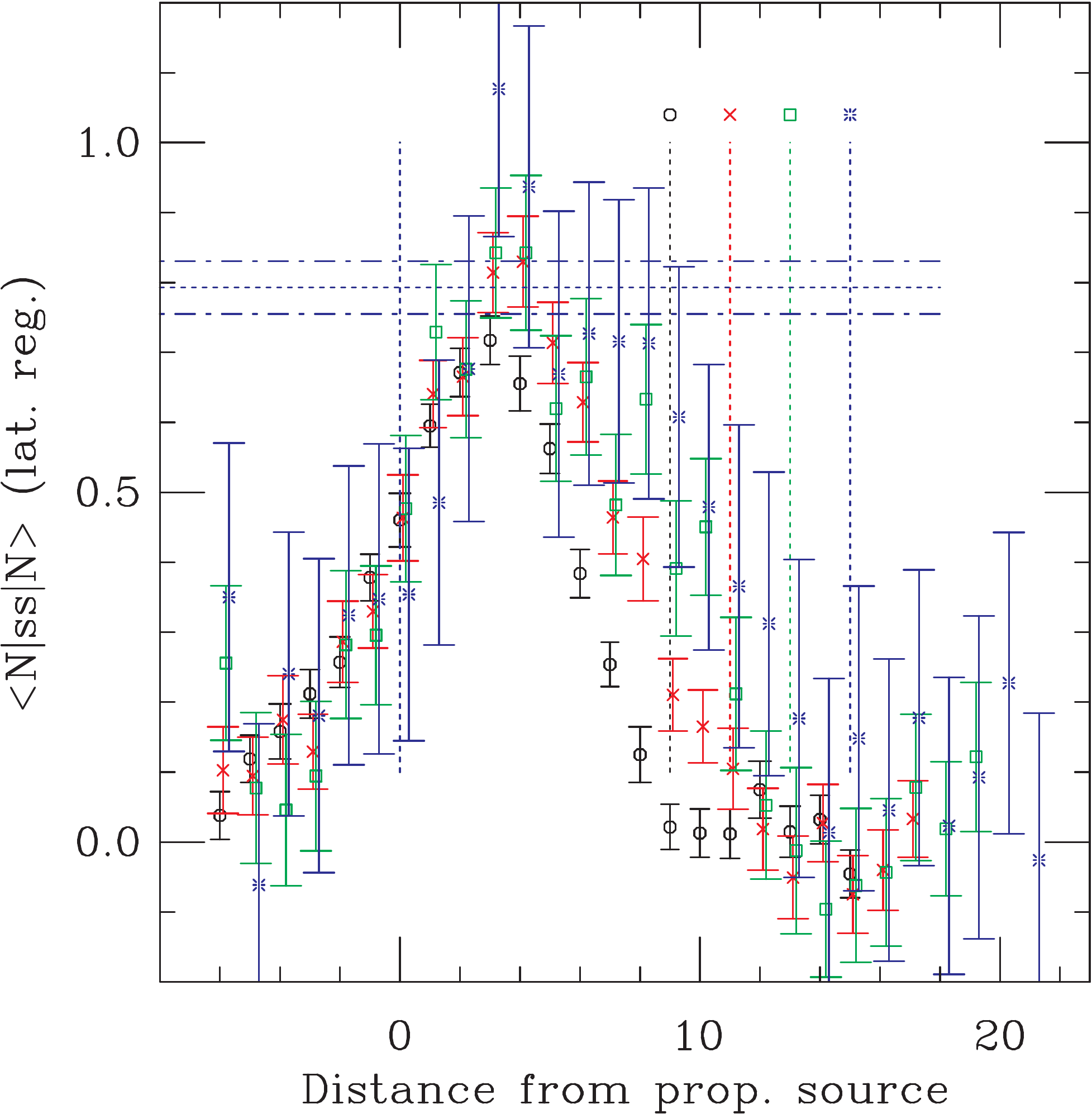}
\caption[$\LL N | \bar s s | N \RR$ using the direct method for multiple propagator lengths]
{The intrinsic strangeness of the nucleon on the MILC Asqtad ensembles using the direct method for
multiple propagator lengths, as a function of $\bar s s$ insertion time $T_1$. Vertical dashed lines represent the source and sink for the various propagators,
with the symbol used to indicate the result using that propagator length indicated above the sink. 
The horizontal lines show the average using the improved hybrid method on these ensembles with the ``conservative'' padding choice.}

\label{fig-direct-manyprops}
\end{figure}

We can also vary the propagator length used. Fig. \ref{fig-direct-manyprops} shows similar curves to Fig. \ref{fig-direct-10} for various length
propagators. As expected, longer propagators have higher statistical error (since the propagators themselves are noisier),
but the height of the plateau itself increases with increasing propagator length. This is not entirely unexpected, since all
of the curves show suppression near the source and sink, and a longer propagator allows for more distance from both. For a
result unpolluted by excited states, we should choose a propagator length $T$ long enough that the result as a function of
$T$ reaches a plateau; for these ensembles, this appears to be the case for $T \approx 12a$. However, these results are far too
noisy to be able to draw much of a conclusion other than that these results appear consistent with those obtained from the improved
hybrid method.

It is also possible and beneficial to use multiple condensate timeslices and multiple propagator lengths to reduce
statistical error. This has a side benefit when applied to a calculation like the present one that uses staggered
fermions. The interpolating operators used to create and annihilate the nucleon
also overlap with states with negative parity whose propagators come with a factor of $(-1)^T$. In particular, the
overlap amplitude with the lowest-lying alternating state is often larger than that of the nucleon itself. In the
hybrid method, this state is included explicitly in the fit forms used to determine $M_N$; however, the
direct method offers no such mechanism. Telltale oscillatory behavior is visible, for instance, in the $T=11$ data
in Fig. \ref{fig-direct-manyprops}.

While these states have masses greater than that of the nucleon and
thus will become irrelevant for condensate measurements taken from
sufficiently long propagators at points sufficiently far from source and sink, their influence can be significant
for the propagator lengths accessible in the current data set. The splitting between these states and the nucleon
is generally less than the splitting with the lowest-lying positive parity excited state, so it will cause
a more substantial pollution of the result. This influence can appear in two places: either
as an alternating trend in the value of $\LL N | \bar s s | N \RR$ as a function of the location $T_1$ of the condensate
measurement, or as a function of the length $T$ of the propagator used. Fig. \ref{fig-direct-manyprops} shows some evidence for this
phenomenon.

To reduce the influence of these alternating states, and to reduce statistical error by making use of more
of the available data, it is useful to consider the condensate on two (or more) adjacent timeslices, and to
consider two adjacent propagator lengths. While this does not ensure that there will be no influence from the
neglected oscillating staggered-fermion state, it should reduce it, while simultaneously improving the statistics.

\begin{figure}[h]
        \includegraphics[width=3.2in]{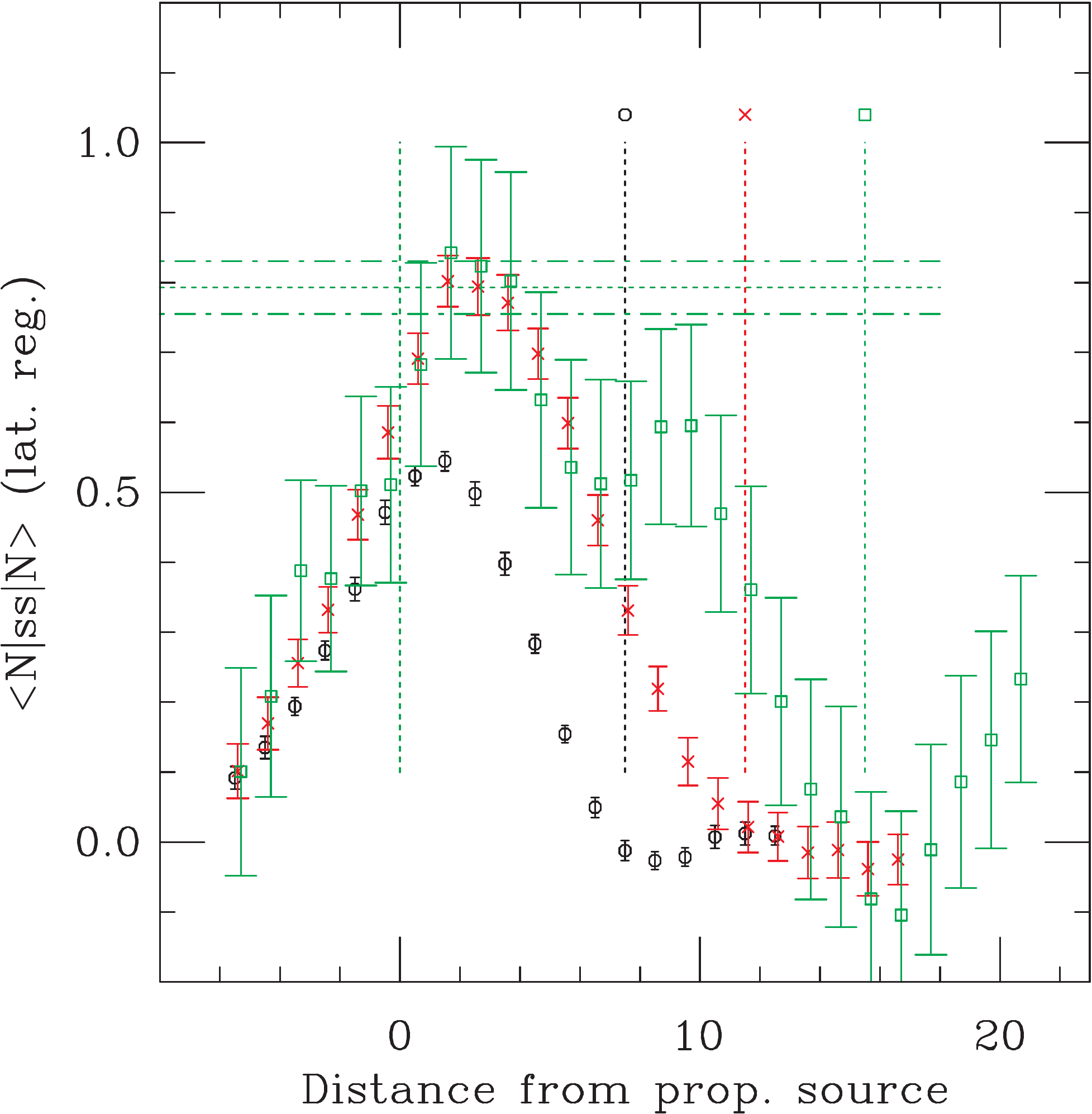}
\caption[$\LL N | \bar s s | N \RR$ using the direct method, averaging over multiple datasets]{Results for the nucleon strangeness $\LL N | \bar s s | N \RR$ from the direct method. 
The results are averaged over
four adjacent condensate measurements at locations centered on the value of $t_1$ shown, and averaged over two adjacent propagator
lengths. The intrinsic strangeness of the nucleon on the MILC Asqtad ensembles using the direct method for
multiple propagator lengths, as a function of $\bar s s$ insertion time $T_1$. The results are averaged over
four adjacent condensate measurements at locations centered on the value of $t_1$ shown, and averaged over two adjacent propagator
lengths. Dotted vertical lines indicate the source and sink locations, with the sink position shown equidistant between the sinks
of the two adjacent propagators that are averaged. Symbols above these dotted lines indicate the plot symbols corresponding to that
propagator length.
The horizontal lines show the average using the improved hybrid method on these ensembles with the ``conservative'' padding choice.}

\label{fig-direct-plot}
\end{figure}

Fig. \ref{fig-direct-plot} shows the resulting values for $\LL N | \bar s s | N \RR$ from the direct method, averaging over two adjacent propagator
lengths and considering the strange quark condensate on multiple timeslices. The result using the improved hybrid method is shown for comparison.
While there is no definitive plateau from which an authoritative value can be taken, the peak values (corresponding to condensate measurements
intermediate between source and sink) agree quite well with those from the improved hybrid method.

These results are not conclusive enough to give a quotable result for $\LL N | \bar s s | N \RR$. However,
there is strong agreement between the direct and improved hybrid methods on this set of ensembles.
As the direct method has been favored by most other recent high-quality calculations of this quantity, its use provides a
useful comparison with their results, and its agreement with the hybrid methods (which are admittedly somewhat convoluted) lends strong support to their validity.

\subsection{Validation from the spectrum method}

As discussed previously, the spectrum method can not be used to produce a reliable calculation of
$\LL N \bar s s | N \RR$ using the MILC Asqtad data. The main difficulty is the tendency to change $\beta$ between
ensembles to keep the lattice spacing, defined {\it via} a Sommer scale and thus dependent on $m_l$, constant. 

However, in the event that several ensembles are available with different quark masses but the same $\beta$ and other lattice parameters ({\em i.e.} the
lattice size and tadpole improvement factor $u_0$),
then some information can be gleaned from calculating
$M_N$ on them. There is only one instance in the Asqtad library where two ensembles have the same $\beta$ and different $m_s$ (and $m_l$). Due
to a mistake in lattice generation, two ensembles were run with $\beta = 7.10$. The first has $a m_l = 0.0093, a m_s = 0.031$, and
the second has $a m_l = 0.0062, a m_s = 0.0186$. These ensembles both have the same dimension ($28^3 \times 96$) and tadpole factor $u_0=0.8785$.

This comparison is made more complicated by the fact that these ensembles differ in both the light and strange quark masses. Thus, we cannot
independently determine $\PAR{M_N}{m_s}$ or $\PAR{M_N}{m_l}$, only a linear combination of the two. Nonetheless, this provides sufficient information
to compare with the larger result from the hybrid method.

Referring to these ensembles as A and B and applying the Feynman-Hellman theorem, we should have

\begin{align}
M_{N,A} - M_{N,B} = (0.0093 - 0.0062) \left(\PAR{M_N}{m_{l,\mathrm{sea}}} + \right.\nonumber\\
\left.\PAR{M_N}{m_{l,\mathrm{val}}}\right) + (0.031 - 0.0186) \PAR{M_N}{m_s}.
\end{align}

Since $\PAR{M_N}{m_{l,\mathrm{sea}}} =  2 \LL N | \bar u u | N \RR_{\mathrm{disc}}$ and $\PAR{M_N}{m_s} =  \LL N | \bar s s | N \RR$, both of which
have been calculated using the improved hybrid method, this suggests a constraint on $M_{N,A} - M_{N,B}$. However, we have no such estimate for
$\PAR{M_N}{m_{l,\mathrm{val}}}$. To eliminate this contribution, we have calculated partially quenched propagators on the $a m_l = 0.0093, a m_s = 0.031$
ensemble with $a m_{\mathrm{val}} = 0.0062$, giving the masses of two nucleons with the same valence quark mass that differ only in their sea quark masses.
Partially quenched propagators were run on only 1020 configurations out of the 1137 equilibriated configurations in the ensemble; the $a m_l = 0.0062, a m_s = 0.0186$
ensemble has 948 configurations.

Referring to the former mass as $M_{N,A'}$, we have

\begin{align}
M_{N,A'} - M_{N,B} &= 2 (0.0093 - 0.0062) \LL N | \bar u u | N \RR_{\mathrm{disc}} \nonumber\\
&+ (0.031 - 0.0186) \LL N \bar s s | N \RR
\end{align}

where the values of $\LL N | \bar u u | N \RR$ and $\LL N \bar s s | N \RR$ should be evaluated not at the physical point but
at quark masses intermediate between ensembles A and B, and the factor of two is due to
the presence of two degenerate light quark flavors in the sea.

\begin{figure}[h]
       \includegraphics[width=3.0in]{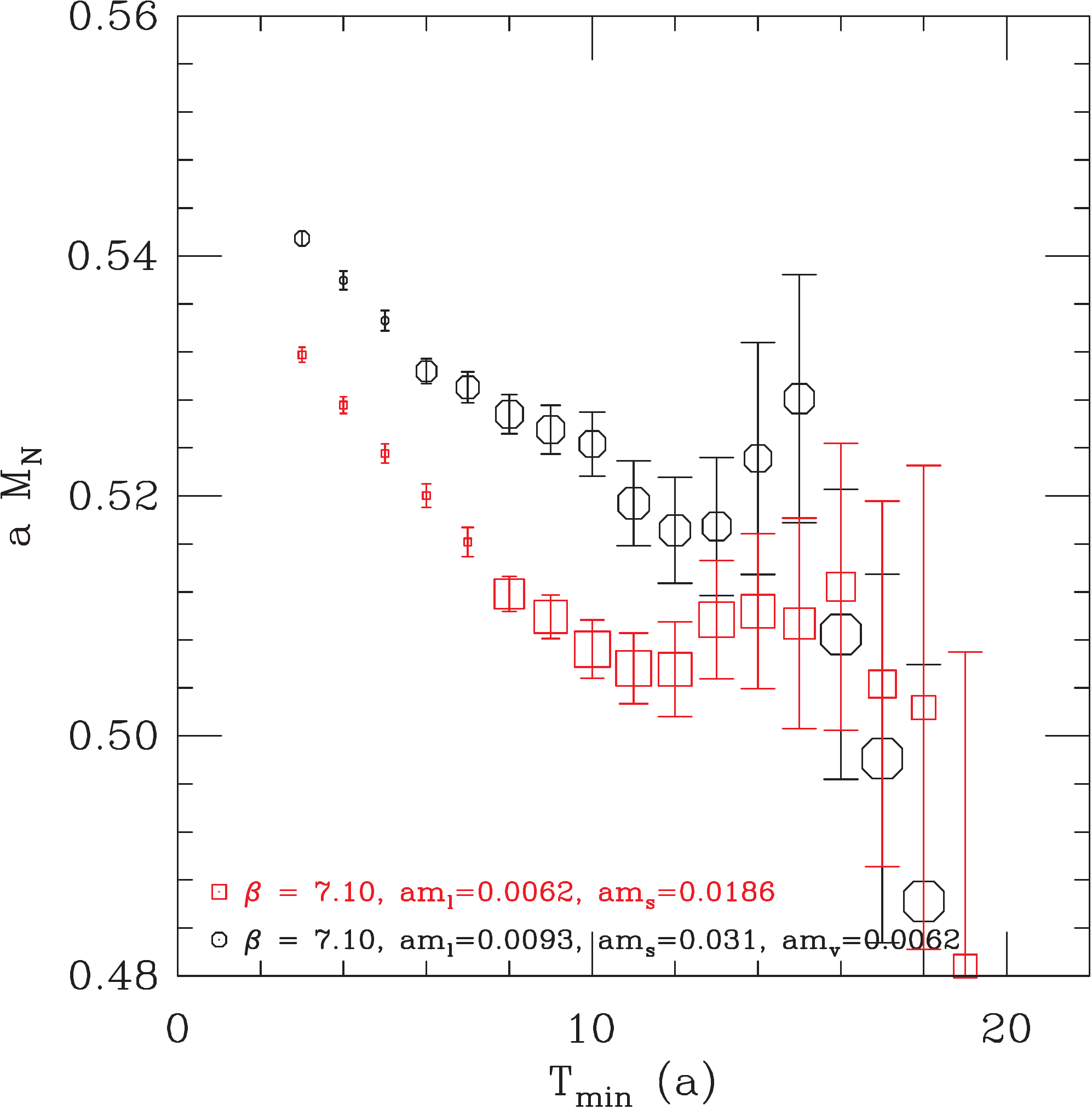}
\caption[The nucleon mass on two ensembles used for a spectrum method calculation]{Results for $M_N$ on the two ensembles with $\beta = 7.10, a m_l = 0.0093, a m_s = 0.031, a m_{\textrm{val}} = 0.0062$ (black octagons) and $\beta = 7.10,
a m_l = 0.0062, a m_s = 0.0186$ (red squares), at varying values of $T_{\mathrm{min}}$, using a two-state fit method.
Symbol size indicates fit quality. Absent data indicate fits that failed to converge or converged to nonsensical values.}
\label{fig-check}
\end{figure}

Fig. \ref{fig-check}
shows the result of a two-state fit to the nucleon propagator on these
two ensembles as a function of the minimum fit distance. 

On both ensembles, both fits produce the expected plateau starting at around $T_{\mathrm{min}} \approx 10$.
$T_{\mathrm{min}}=11$
represents the best compromise between statistical error and possible excited-state pollution for determining the difference in $M_N$; this gives
the values $M_{N,A'} = 0.519(4)$ for the $a m_l = 0.0093, a m_{\mathrm{val}} = 0.0062$ ensemble
and $M_{N,B} = 0.506(3)$ for the $a m_l = 0.0062$ ensemble. Combining the errors in quadrature, we get $M_{N,A'} - M_{N,B} = 0.013(5)$

To test this against the hybrid method, we calculate $(0.0093 - 0.0062) \LL N | \bar u u | N \RR + (0.031 - 0.0186) \LL N \bar s s | N \RR$
which should give a similar result. First, it is simplest to average the values of $\LL N | \bar u u | N \RR$ and $\LL N | \bar s s | N \RR$ on these
two ensembles. Results from the hybrid method with $T_{\mathrm{min}}=7$ are given in Table \ref{table-spec-check}.

\begin{table}
\centering\begin{tabular}{|c | c c | }
\hline
Ensemble                         & $\LL N | \bar s s | N \RR$ & $\LL N | \bar u u | N \RR$ \\
\hline
$a m_l = 0.0093, a m_s = 0.031$ & 1.09(18) & 1.75(28) \\
$a m_l = 0.0062, a m_s = 0.0186$ & 0.77(22) & 0.53(36) \\
\hline
Weighted average & 0.96(14) & 1.29(22) \\
\hline
\end{tabular}
\caption[Results for heavy and light quark content of the nucleon on two ensembles used for a direct method calculation]{Results for heavy and light quark content of the nucleon on $\beta = 7.10, a m_l = 0.0093, a m_s = 0.031$ and $\beta = 7.10,
a m_l = 0.0062, a m_s = 0.0186$ ensembles, and a weighted average.}
\label{table-spec-check}
\end{table}

Note that ensemble B has $\LL N | \bar s s | N \RR < \LL N | \bar u u | N \RR$, which is unexpected;
this difference, however, is not statistically significant ($\LL N | \bar s s | N \RR - \LL N | \bar u u | N \RR = 0.24(28)$, with the
error bar computed in the proper way {\it via} jackknife).

These values give $(0.0093 - 0.0062) \LL N | \bar u u | N \RR + (0.031 - 0.0186) \LL N \bar s s | N \RR = 0.011(4)$, in reasonable agreement with the
$0.013(5)$ estimated from the difference of the nucleon masses. 
This agreement between the hybrid and spectrum methods on these two ensembles lends support to the validity of the former.

\section{$\LL N | \bar s s | N \RR$ from the MILC HISQ data}

The hybrid method can also be applied to the newer HISQ data with no significant modification, although none of the
measurements required to apply the improved hybrid method are available. 
Several features of the HISQ dataset make it somewhat less suitable for analysis. There is a more limited range of light quark
masses available, ranging from the physical value to $0.2 m_s$. This greatly reduces the ``lever arm'' available to determine
the slope of the chiral extrapolation compared to the Asqtad ensembles, where each nominal lattice spacing had runs with
$m_l = 0.4 m_s$, and one ensemble with $a \approx 0.09\ \textrm{fm}$ has $m_l = m_s$ (three degenerate heavy quarks).
The HISQ ensembles are all limited to 1000 equilibriated gauge configurations, and the gauge generation program is still
ongoing, limiting the statistics available.

Nonetheless, essentially the same analysis can be carried out. The form of the chiral and continuum
extrapolations will be unchanged, although the value of the constant controlling the continuum
extrapolation should be smaller since taste breaking, the dominant discretization error in the Asqtad action, is reduced roughly
by two-thirds in HISQ. Since the continuum extrapolation does not have much of an effect, we keep the prior with a width corresponding to a
10\% discretization effect from $a = 0.12$ fm to the continuum; while this effect should be reduced in HISQ, we have no basis for estimating
to what extent that is the case. The values of $Z_m$ to convert to the $\overline{\mathrm {MS}}$ (2 GeV)
regularization are, to sufficient accuracy, the same for HISQ and Asqtad.

The same minimum distance to consider for the nucleon fits in order to avoid excited state pollution is also
still sensible; nothing in the physics which controls excited state pollution is affected by the further improved fermion action. 

\begin{figure*}
\hspace{-.5in}
        \includegraphics[width=2.8in]{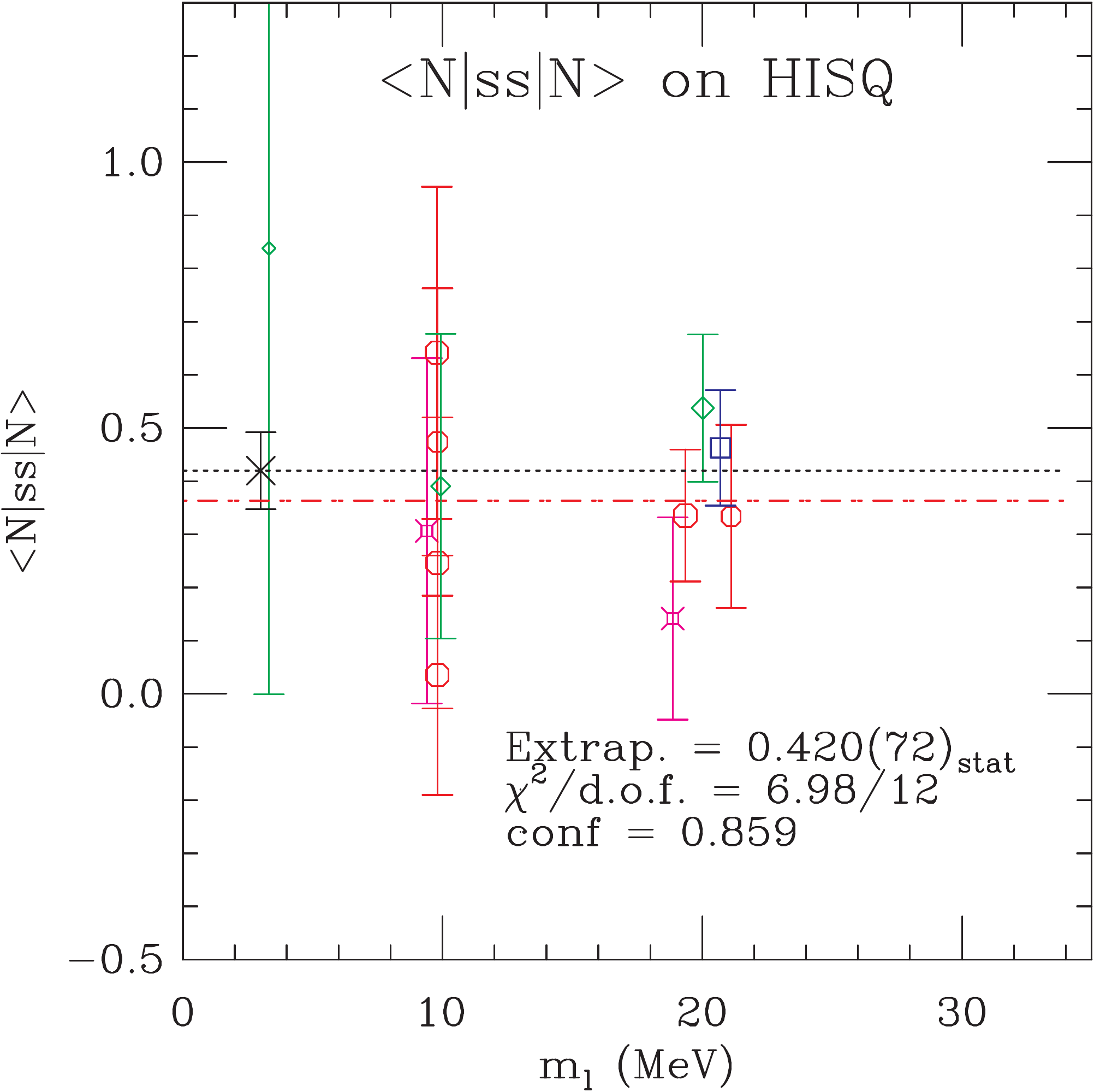}
        \hspace{.5in}
        \includegraphics[width=2.8in]{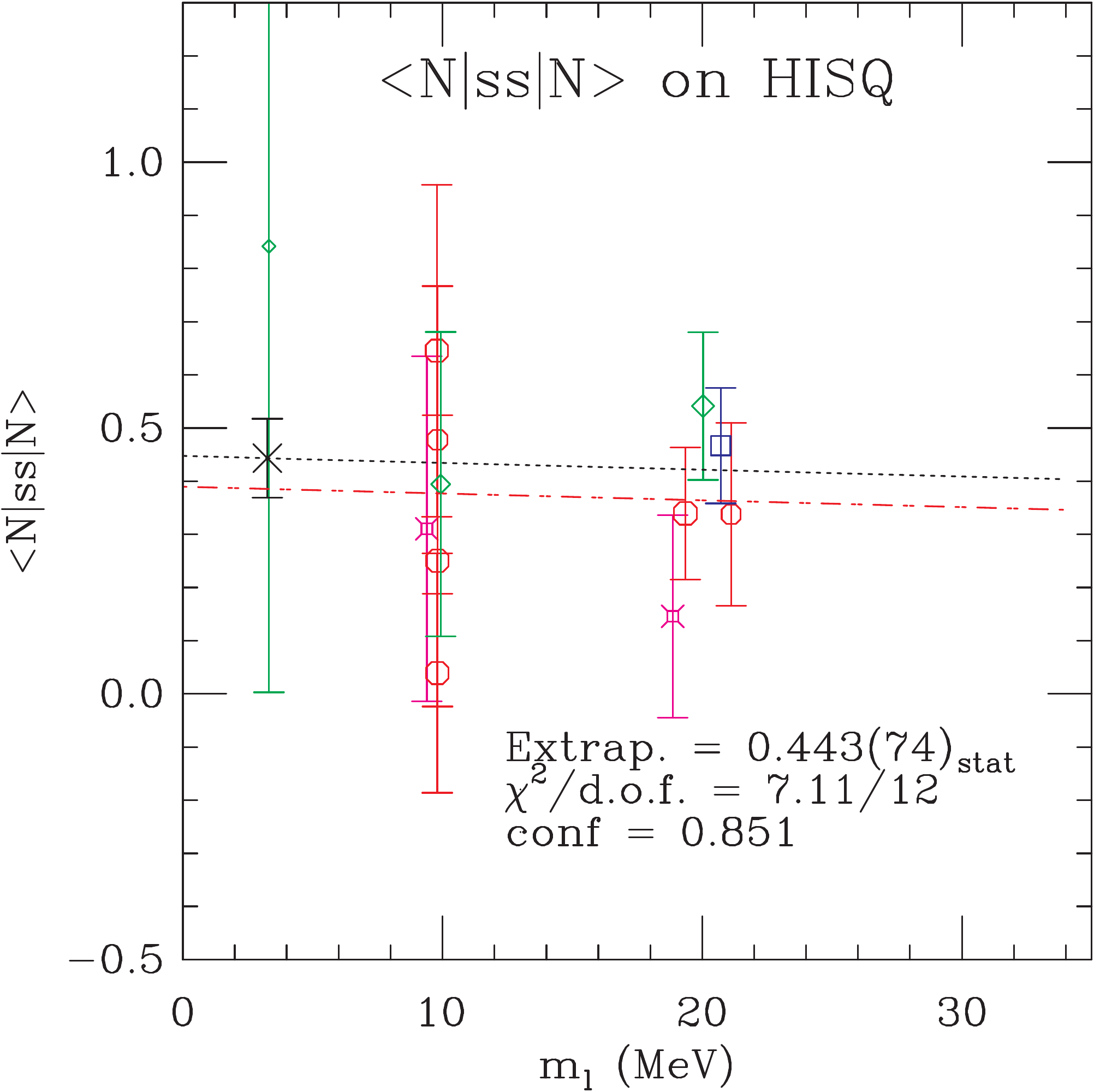}
\caption[$\LL N | \bar s s | N \RR$ on the MILC HISQ ensembles using the hybrid method]
{The intrinsic strangeness of the nucleon on the MILC HISQ ensembles, using the hybrid method. The left pane shows
a constant fit; the right pane shows a linear chiral fit, with the slope constrained using a Bayesian prior from the Asqtad fit.
Ensembles with $a \approx (0.15, 0.12, 0.09, 0.06)$ fm are shown as
violet fancy squares, red octagons, green diamonds, and blue squares, respectively. Symbol area is proportional to the number of gauge
configurations in the ensemble. The fit at the continuum is shown as a black dotted line,
while the fit evaluated at $a=0.12$ fm is shown as a red dashed line. The fit evaluated at the continuum is shown as a black cross, and the
quoted error includes the uncertainty in the continuum extrapolation with the prior as discussed in the text.}
\label{fig-hisq-strangeness}
\end{figure*}

Applying the same analysis described in detail previously, we obtain nonsensical results due to the lack of sufficient lever arm to constrain
the slope of the chiral extrapolation, due to the lack of ensembles with heavier light quarks. While we may omit the $m_l$ dependence from the fit, a somewhat
more sophisticated approach is to constrain the slope of the chiral fit. We use a Gaussian Bayesian prior whose central value is taken from the Asqtad fit and whose
width is equal to the error of the Asqtad slope. The constant fit and the fit with a constrained slope are shown in Fig. \ref{fig-hisq-strangeness}.

These two fits are very similar; we consider the fit with the constrained slope to be more physical.

The systematic errors should be similar to those in the Asqtad case. Since the HISQ ensembles have larger physical volumes, we use 2\% as an estimate of the systematic
error from finite-volume effects. The error budget is summarized in \ref{table-hisq-error-budget}.

\begin{table}
\centering
  \begin{tabular} { | c | c | }
  \hline
  Source & Error\\
  \hline
  Statistical & 0.08 \\
  Higher order $\chi$PT & 0.03 \\
  Excited states & 0.02 \\
  Finite volume & 0.01 \\
  Renormalization & 0.03 \\
  \hline
  \end{tabular}
  \caption[Error budget for $\LL N | \bar s s | N \RR$ using the hybrid method on the HISQ ensembles]{Error budget for the measurement of $\LL N | \bar s s | N \RR$ using the unimproved hybrid method on the HISQ data.}
  \label{table-hisq-error-budget}
\end{table}

Thus, our result extrapolated to the physical point is $\LL N | \bar s s | N \RR=0.44(8)_{\rm{stat}}(5)_{\rm{sys}}$.

\section{$\LL N | \bar c c | N \RR$ from the MILC HISQ data}

The intrinsic charm of the nucleon can also be measured on the lattice. 
This is interesting both in its own right and because it can be compared with the perturbative prediction. 
Since the lattice calculation directly measures $\LL N | \bar c c | N \RR$, we convert Kryjevski's perturbative
computation\cite{KRYJEVSKI03} to this form. Using $x_{uds}=0.14$ and
$m_c = 1.2$ GeV, the perturbative prediction is $\LL N | \bar c c | N \RR = 0.057$.
Due to the smaller absolute magnitude of the intrinsic charm, and the lower statistics of the HISQ ensembles used to determine it, it is even more difficult to
extract than the nucleon strangeness. 

\subsection{The lattice result}

\begin{figure*}[t]
        \includegraphics[width=2.9in]{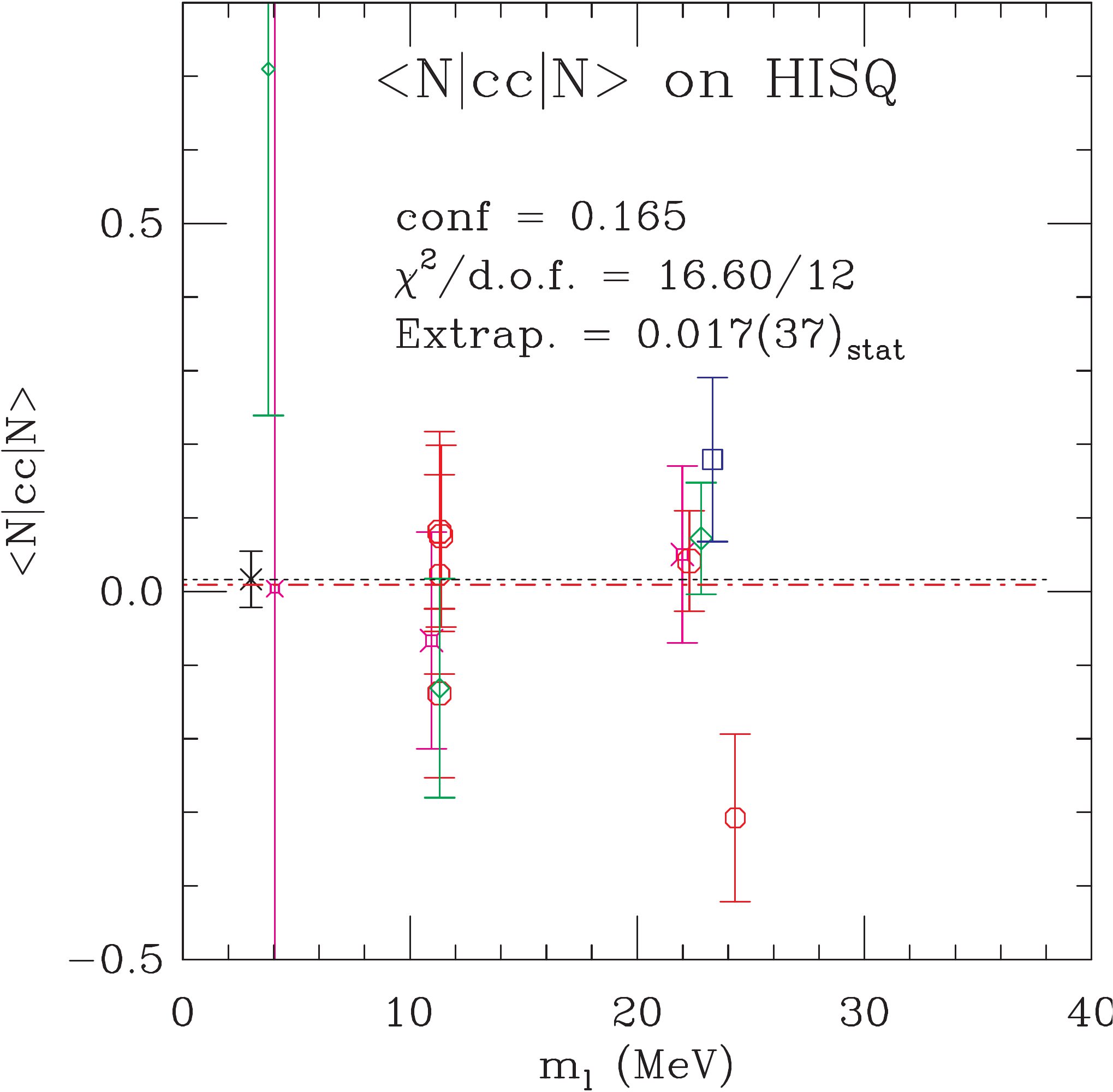}\hspace{0.9in}
        \includegraphics[width=2.8in]{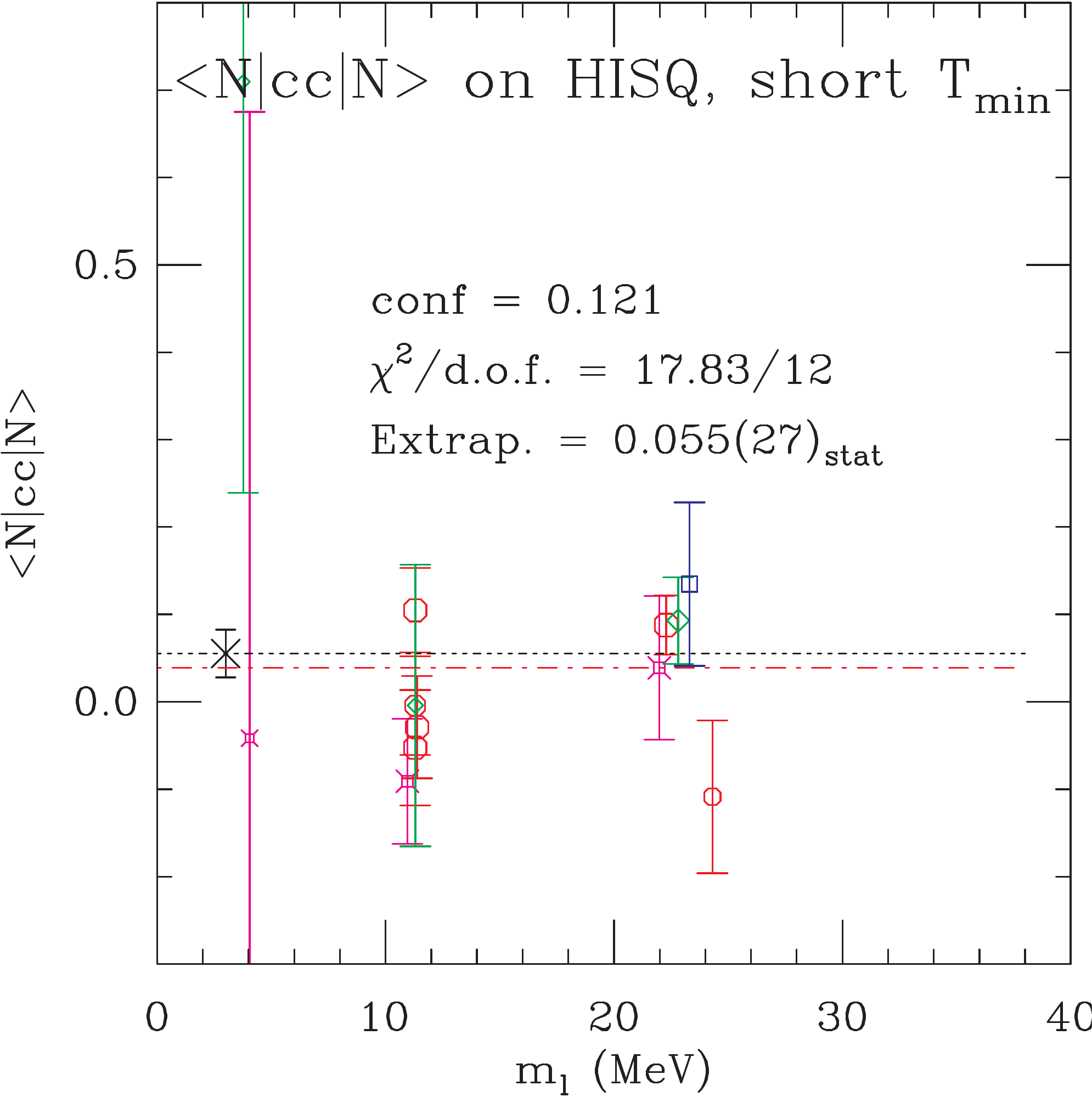}

\caption[The intrinsic charm content of the nucleon on the MILC HISQ ensembles using the hybrid method]
{The intrinsic charm content of the nucleon on the MILC HISQ ensembles using the hybrid method. The left
pane shows the result using the larger minimum propagator distances $t_{\mathrm{min}}$ used for
the strangeness calculation; the right pane shows the result using the smaller minimum propagator
distances discussed in the text. Ensembles with $a \approx (0.15, 0.12, 0.09, 0.06)$ fm are shown as
violet fancy squares, red octagons, green diamonds, and blue squares, respectively (color online).
Symbol area is proportional to the number of gauge configurations in the ensemble. The fit at the
continuum is shown as a black dotted line, while the fit evaluated at $a=0.12$ fm is shown as a red
dashed line. The fit evaluated at the continuum is shown as a black cross, and the
quoted error includes the uncertainty in the continuum extrapolation with the prior as discussed in the text.}
\label{fig-hisq-charmness}

\end{figure*}

The application of the hybrid method to the nucleon intrinsic charm proceeds identically to the nucleon strangeness,
except that the charm quark condensate is used. Pollution due to excited states should have a similar
impact on the intrinsic charm and the intrinsic strangeness, so for a first analysis we use the same set of
minimum fit distances for $M_N$ as before.
The results on the various HISQ ensembles are shown
in Fig. \ref{fig-hisq-charmness}, along with the fit. The results are broadly consistent with the
perturbative prediction, but indistinguishable from zero due to the large fractional statistical errors.
We thus do not attempt a chiral extrapolation. For the continuum extrapolation, we use the same procedure
as for the nucleon strangeness, by adding a term proportional to $a^2$ to the fit. Since the coefficient
of such a term would be extremely poorly determined given the high statistical errors, we use the same procedure as
before to constrain it (see Sec. \ref{sec-continuum}) by the imposition of a Bayesian prior with a width
corresponding to a 10\% effect between $a = 0.12$ fm and the continuum. (Note however that as
the 10\% figure was obtained from Asqtad lattices, smaller lattice discretization effects might be expected
on HISQ if these effects stem from taste-breaking interactions.) This difference is not terribly
meaningful, of course, given the size of the statistical errors involved.  However, in this case the fit
value of the nucleon intrinsic charm is very nearly zero, so the allowance for a 10\% effect would lead
to an artificially tight prior.
Thus, we impose a prior that allows for an 0.02 shift between $a = 0.12$ fm and the continuum,
which is roughly 1/3 the perturbative prediction.
(In practice, the width of this prior has essentially no effect on such a noisy fit.)

Recall that the choice of the minimum propagator length $t_{\mathrm{min}}$ used to fit the nucleon propagator was made
to provide the best available balance between the statistical error (reduced for lower values of $t_{\mathrm{min}}$)
and the potential for systematic error due to excited-state pollution (larger at lower values of $t_{\mathrm{min}}$).
The balance between these two errors struck for the nucleon strangeness was appropriate for a quantity with
smaller statistical errors, but when dealing with the nucleon charm the impact of additional excited state pollution is
not as meaningful when dealing with a quantity with such a large statistical error. Additionally, per the
perturbative argument \cite{SHIFMAN78,KRYJEVSKI03},
excited-state pollution should matter less when dealing with heavy quarks, since the effect of altering the mass of a quark
well above the scale $\Lambda_{\textrm{QCD}}$ affects all low-energy quantities ({\em i.e.} the masses of
the nucleon and its excited states) in the same way, interpreted as an overall rescaling of the lattice. 
This is not strictly true, of course, since the charm quark is not that much greater in mass
than $\Lambda_{\textrm{QCD}}$; otherwise, the inclusion of dynamical charm in simulations of
low-energy quantities would be meaningless, since by the same argument all they would do is change the lattice spacing!
Nonetheless, it suggests that the effect due to excited state pollution is less for the intrinsic charm
than for the strangeness.

It is thus appropriate to use smaller minimum distances in evaluating the
nucleon intrinsic charm. As before, we should choose a minimum distance that is relatively constant in
physical units across lattice spacings. No methodical evaluation of the systematic errors due to excited
state pollution can really be made with these data, so we choose (in a rather ad-hoc way) to use minimum
distances that are roughly $2/3$ those used for the strangeness. (A somewhat artificial lower bound on the
minimum distances used, especially for the coarser ensembles, is provided by fitter convergence; if
there is too much excited state pollution, the nucleon mass fits may behave unpredictably.) We thus choose
$t_{\mathrm{min}}=(3a, 3a, 5a, 7a)$ for the $a \approx (0.15, 0,12, 0.09, 0.06)$ fm ensembles, respectively.
 
Using the same minimum distances that were used in for
$\LL N | s \bar s | N \RR$, ($10a$ for $a \approx 0.06$ fm, $7a$ for $a \approx 0.09$ fm, $5a$ for
 $a \approx 0.12$ fm, and $4a$ for $a \approx 0.15$ fm),
we would find that $\LL N | \bar c c | N \RR = 0.017(37)_{\mathrm{stat}}$.
Using the smaller minimum distances discussed above, we obtain $\LL N | \bar c c | N \RR = 0.056(27)_{\mathrm{stat}}$ if the
coarsest ensembles are included, and $\LL N | \bar c c | N \RR = 0.054(27)_{\mathrm{stat}}$ if they are
omitted. This result has, as expected, lower statistical error, in agreement with the perturbative
prediction.\cite{KRYJEVSKI03}.
We make no estimates of systematic errors because of the large size of the 
statistical uncertainty. These values are broadly consistent with the perturbative predictions
given by Kryjevski\cite{KRYJEVSKI03}.

\section{Conclusions}

Early lattice work on the nucleon strangeness was prompted in part by early $\chi$PT calculations
suggesting that its value might be large\cite{NELSON87}, and thus provide
a significant enhancement to the WIMP-on-baryon scattering cross-section. While there is 
no reason to expect the perturbative method in Refs.\cite{SHIFMAN78,KRYJEVSKI03} to apply
accurately to the strange quark, that result can be used to
set a natural scale for $\PAR{M_N}{m_s} \approx \frac{2}{29}\frac{M_N}{m_s} \approx 0.7$.
Early lattice calculations suffered from large uncontrolled systematics,
leading to wildly varying estimates for the nucleon strangeness, some of which were large.
This work, as well as our previous work on the nucleon strangeness
in Ref. \cite{OURPRL} and many other recent calculations\cite{OHKI08,JLQCD09,JLQCD10_POS,ENGELHARDT2010}, 
however, all conclude that the nucleon strangeness is approximately at its natural scale, substantially
smaller than the early work suggested that it might be. While the uncertainty in the value
of $\LL N | \bar s s | N \RR$ is still large compared to other
quantities, it is no longer a dominant contribution to the uncertainty in dark matter scattering amplitudes;
from the perspective of dark matter cosmology,
the problem of the strangeness of the nucleon presented in Refs. \cite{BALTZ06,ELLIS08} has been solved.

We apply the ``hybrid method'', outlined in Sec. \ref{sec-hybrid}, to the large library of improved
staggered gauge configurations (roughly 26000 Asqtad and 14000 HISQ)
generated by the MILC Collaboration.
Improved staggered fermions are well-suited to this project, since they are very fast, allowing for large
gauge ensembles with which to beat down the inherently noisy disconnected diagrams involved here,
and they preserve a remnant chiral symmetry, allowing for a straightforward application of the
Feynman-Hellman theorem without the concerns about additive renormalization or operator mixing
which plague Wilson-based computations of this quantity, as discussed in Ref. \cite{MICHAEL01}. 

We present an improvement on this method which, due to historical averaging over sources and timeslices when measuring $\bar q q$ and the nucleon propagator, required
repeating the lattice measurements of these quantities; this was thus only performed on the coarser lattices in the Asqtad library (some 14000 configurations out of 26000 total)
to economize on computer time. 
We determine that the condensate should be considered out to a distance of approximately 0.7 fm from the source and sink of the nucleon
propagator. This results in an approximately 40-50\% decrease in the statistical error.

Using the improved hybrid method where the needed measurements are available, we conclude that {\bf $\LL N | \bar s s | N \RR = 0.637(55)_{\rm{stat}}(74)_{\rm{sys}}$}
after continuum and chiral extrapolation.
The leading uncertainties are statistical, mainly in the continuum extrapolation, and in the systematic error due to excited state contamination.
Since smaller statistical errors allow the use of larger minimum distances (to reduce excited state errors) or allow the justification of smaller systematic error estimates
for the minimum distances chosen, improvements in statistics will also lead to improvements in this systematic error.
Availability of the needed measurements to apply the improved hybrid method on more of the finer ensembles would help with this, as well as 
helping to reduce the large contribution to the statistical error from the continuum extrapolation.

This method can also be used to calculate the nucleon strangeness using the HISQ fermion action. Due to the lack of large-$m_l$ ensembles in the HISQ program, we constrain
the slope of the chiral extrapolation using a Gaussian Bayesian prior whose central value and width are taken from the Asqtad fit. We obtain
$\LL N | \bar s s | N \RR = 0.437(8)_{\rm{stat}}(5)_{\rm{sys}}$.

The same methods can be used to compute the intrinsic charm of the nucleon. This provides a connection to perturbative QCD.
A simple argument involving the running of the QCD coupling constant and the fact that the low-energy QCD scale is set by the running of $g$
can be used to estimate $\LL N | \bar c c | N \RR$\cite{SHIFMAN78, NELSON87};
that result has been improved to four-loop order by Kryjevski\cite{KRYJEVSKI03}, who calculates $\LL N | \bar c c | N \RR = 0.058$.

Using the MILC HISQ ensembles with dynamical charm, we apply similar methods as with the intrinsic strangeness.
Excited-state pollution is less of a problem for this quantity; by Kryjevski's perturbative argument,
the effect of altering the charm quark mass mostly just rescales the lattice, so the intrinsic charm of the
excited states of the nucleon interpolating operator should be similar to that of the ground state.
Moreover, we are dealing with fractional statistical errors that are about an order of magnitude larger.
Thus, we choose smaller minimum propagator fit distances $T_{\rm{min}}$ to improve the statistics.
Similarly, because the overall statistical error is so high, we do not attempt a chiral extrapolation.

Doing the continuum fit in the same manner as before, we obtain
$\LL N | \bar c c | N \RR = 0.056(27)_{\mathrm{stat}}$, essentially the same as the perturbative
result albeit with large fractional error due to the lower value. As this error is dominated by
statistics, we do not present a systematic error budget.  This result can be improved by the
simple availability of more HISQ data, as the MILC HISQ lattice generation project progresses.

Taken together, these results for the scalar strange and charm quark content of the nucleon, similar
results for the strange quark content from other groups (Fig.~\ref{fig-history}), 
and the perturbative calculation,
lead to the amusing conclusion that the scattering of a low-momentum Higgs particle (as in dark
matter interactions in the MSSM) is not dominated by the strange quarks in the nucleon, but instead
receives more or less equal contributions from all heavy quarks.

\section*{Acknowledgements}This work was supported by the Department of Energy grants DE-FG02-04ER-41298 and DE-FG02-95ER-40907.
Computer time was provided through
     the National Energy Resources
    Supercomputing Center (NERSC), which is supported by the Office of
    Science of the U.S. Department of Energy under Contract No.
    DE-AC02-05CH11231.
    Computer resources were also provided by
    the USQCD Collaboration, the Argonne Leadership Computing Facility,
    and the Los Alamos National Laboratory Computing Center.
    which are funded by the U.S. Department of Energy.
    Computer resources were also provided by the National Science Foundation's NRAC, Teragrid and XSEDE programs,
    including the National Center for Supercomputing Applications (NCSA),  
    the Texas Advanced Computing Center (TACC),   
    the Pittsburgh Computer Center (PSC),   
    the National Institute for Computational Sciences (NICS),
    and the San Diego Supercomputer Center (SDSC).
    Computer time at the National Center for Atmospheric Research 
    was provided by NSF MRI Grant CNS-0421498, NSF MRI Grant
    CNS-0420873, NSF MRI Grant CNS-0420985, NSF sponsorship of the
    National Center for Atmospheric Research, the University of Colorado,
    and a grant from the IBM Shared University Research (SUR) program.
We thank David Kaplan, Ulf Meissner, Michael Engelhardt, and our MILC collaboration colleagues for helpful
discussions and suggestions.

\end{document}